\documentclass[aps]{revtex4-1}
\usepackage{amsmath,amssymb}
\usepackage{color,graphicx}
\usepackage{slashbox}
\newcommand{\be}{\begin{equation}}
\newcommand{\ee}{\end{equation}}
\begin{document}
\title{Solving fractional Schr\"{o}dinger-type spectral problems: Cauchy oscillator  and
  Cauchy  well}
\author{Mariusz  \.{Z}aba and   Piotr Garbaczewski}
\affiliation{Institute of Physics, University of Opole, 45-052
Opole, Poland}
\date{\today }
\begin{abstract}
This paper is a direct offspring of Ref. \cite{GS} where basic tenets of the nonlocally induced random
and quantum dynamics were  analyzed. A number of  mentions  was maid with respect to  various inconsistencies
and faulty statements  omnipresent in the literature devoted to  so-called  fractional quantum  mechanics spectral problems.
Presently,  we  give  a decisive  computer-assisted  proof, for an exemplary  finite and ultimately infinite  Cauchy  well  problem,
 that   spectral solutions proposed so far were plainly wrong. As a constructive input, we  provide an explicit  spectral solution of
 the finite Cauchy well. The infinite well emerges as a  limiting case  in a  sequence of
 deepening  finite wells.    The employed numerical methodology  (algorithm based on the Strang splitting method)
 has been tested  for an  exemplary  Cauchy oscillator problem, whose analytic solution is available.
 An impact of the inherent  spatial   nonlocality of motion generators upon  computer-assisted   outcomes  (potentially  defective, in view of  various cutoffs),
 i.e.  detailed  eigenvalues and shapes of eigenfunctions,  has been analyzed.
 \end{abstract}
\maketitle

\section{Introduction}

A fully fledged theory of quantum dynamical patterns of behavior that
are nonlocally induced has been  carefully analyzed in  Ref. \cite{GS}.
The pertinent motion scenario stems from generalizing  the standard Laplacian-based
framework of the Schr\"{o}dinger picture quantum evolution to that employing nonlocal
(pseudo-differential) operators. Except for so-called relativistic Hamiltonians
(c.f.  the spinless Salpeter equation, with or without external local potentials \cite{GS,R1,Li}), generic
nonlocal energy generators are not supported by a classical mechanical  intuition of
massive particles in motion.

 A particular  mass zero version  (Cauchy generator)
 of the relativistic Hamiltonian may be linked with the photon wave mechanics  and Maxwell fields \cite{GS}.
Likewise, other fractional generators  seem to  have more in common  with  fields than with
 particles, regardless of their classical or quantum connotations.

The   standard  unitary  quantum  dynamics   $\exp(-iHt/\hbar )$  and the  Schr\"{o}dinger
semigroup-driven random motion  $\exp (-tH/\hbar )$  are examples of dual evolution scenarios that
may be mapped among each other  by means of a suitable   analytic
continuation in time (here e.g. $it \rightarrow t$ for times $t\geq 0$),
   which is a quantum mechanical offspring of the  Euclidean quantum theory methods.
 Both  types of motion   share in  common a local  Hamiltonian operator $H$.
 Its  spectral resolution is known to determine simultaneously:   (i) transition
probability amplitudes of the Schr\"{o}dinger picture  quantum
motion in $L^2(R^n)$  and  (ii) transition probability densities of  a
space-time homogeneous  diffusion  process in $R^n$, with $n\geq 1$.

Within the general theory of so-called infinitely divisible probability laws  the
familiar Laplacian (Wiener noise or Brownian motion generator) is known to be one isolated
member of a surprisingly rich family of non-Gaussian L´evy noise generators. All of them stem
from the fundamental L\'{e}vy-Khintchine formula, and  refer to  (heavy-tailed)   probability distributions of
spatial jumps and the resultant jump-type Markov processes.

The emergent L\'{e}vy generators are manifestly nonlocal (pseudo-differential) operators and
 give rise (via a  canonical quantization procedure, \cite{GS}) to L\'{e}vy-Schr\"{o}dinger semigroups
and affiliated nonlocally-induced jump-type processes.
 The dual (Euclidean)  image of such semigroups comprises unitary dynamics scenarios which  give rise to
 a   nonlocal   quantum behavior.\\

{\bf Remark  1:}  The  canonical quantization is  introduced  indirectly,  by means of the
 most primitive  ansatz, whose core lies
in choosing   the Hilbert space $L^2(R^n)$ as an arena for our investigations. From the start we have the
Fourier transformation realized as a unitary operation in this space and a canonical quantization
input as an obvious  consequence. The above mentioned L\'{e}vy-Khintchine formula,  actually   derives
from a Fourier transform of a symmetric probability density function. A variety of symmetric
probability laws for random noise is classified by means of a characteristic function which is
an exponent ($\eta (p)$ of the $(2\pi )^{n/2}$-multiplied Fourier transform of that probability density function (pdf)
$\int   d^nx\, \rho (x) \exp(\pm ipx)= \exp[\eta (p)]$.
The naive   (note that $\hbar \equiv 1$) canonical quantization step $p  \rightarrow \hat{p} = -i\nabla $, while
 executed  upon  the characteristic function, induces   random  jump-type processes  that are
 driven by L\'{e}vy-Schr\"{o}dinger semigroups $ \exp[ t\eta (\hat{p})]$.
 Their dual partners actually are the unitary evolution operators $ \exp[it \eta (\hat{p})]$
of interest and are a subject  of further discussion, specifically after incorporating
locally defined  confining potentials and/or  boundary data.
 \\

{\bf Remark 2:} Redefining the characteristic exponent as $\eta (p) = - F(p)$, we can
actually classify a subclass of  probability laws (and emergent generators) of interest.
 Those are: (i) symmetric stable laws that correspond to $F_{\mu }(p) = |p|^{\mu } $, with
 $\mu \in (0,2)$ and (ii) relativistic probability law inferred from   $F^m(p) =
 \sqrt{p^2+ m^2} - m$, $m>0$ which is  a rescaled (dimensionless)
  version of a classical relativistic  Hamiltonian  $\sqrt{  m^2c^4 + c^2p^2} - mc^2$,
  where $c$ stands for the velocity of light ($c\equiv 1$ is used throughout the paper.  We note that $F_{1}(p) = |p|$  determines
  the Cauchy probability law and gives rise to  Cauchy  operator, here  denoted
   $|\nabla | = (- \Delta )^{1/2}$. Clearly $F_1(p)=F^0(p)$.
\\

 With no explicit mention of the affiliated stochastic formalism (e.g.  L\'{e}vy jump-type processes),
 nonlocal Hamiltonians  of the form  $\sim (- \Delta )^{\mu /2}$ are instrumental in
  so-called fractional quantum mechanics \cite{Laskin}-\cite{D}. Its   formalism, as
 developed so far,  is  unfortunately not  free  from  inconsistencies  and
 disputable  statements. Specifically this happens with regard to spectral problems, where perturbations by  confining
(local) potentials  or   a priori imposed   boundary data,   yield  a discrete  energy spectrum together with related eigenfunctions.
Their functional properties   (analytic or numerical analysis of shapes) have  not  been unambigously settled.

In this connection, the need for a careful account of the nonlocal character of the motion generator
has been clearly exposed in Refs.  \cite{J,YL}.   Its neglect in solution procedures for   potentially  simplest
 infinite well problem has  led to erroneous formulas for both eigenvalues and eigenvectors  of the
 generator. Those  have been repeatedly reproduced  in the literature, \cite{Laskin}-\cite{D}.
Currently, the status of undoubtful  relevance have    approximate statements (various   estimates)  pertaining
to the asymptotic behavior of  eigenfunctions   at the well boundaries and estimates, of varied degree of accuracy,
 of the   eigenvalues, c.f. \cite{Z} and \cite{BK}-\cite{K}.

We note in passing  that even in the fully local (Laplacian-based)  case,
one should not   hastily  ignore the exterior of the entrapping enclosure,  if
the  canonical quantization  is to be reconciled with the existence of  impenetrable
barriers  and the infinite well problem  in particular, \cite{GK}.

Historically the  nonlocality   of relativistic generators (likewise, that of  fractional ones)
 has been  considered as a nuisance, sometimes elevated to the status of  a devastating  defect of
 the theory and  a "good  reason"  for its abandoning. Thence   not as  a property worth  exploitation on its own.
However, we can proceed  otherwise.  If one  takes  nonlocal Hamiltonian-type   operators  seriously,  as
  a conceptual broadening of current quantum paradigms,  it is of interest
  to analyze  solutions of  the resultant  nonlocal  Schr\"{o}dinger-type equations. Of particular importance are
  solutions of related spectral problems   in  confining  regimes, which are  set either by   locally defined  external  potentials
  or by  confining boundary data.

If an   analytic  solution of  the "normal"  Laplacian-based
  Schr\"{o}dinger eigenvalue problem  is not in the reach,  a recourse   to the
    imaginary time propagation technique  (to evolve the system in "imaginary time", to employ "diffusion algorithms") is a standard routine
 \cite{BBC}-\cite{Chin}, see also \cite{A}.
  There exist a plethora of methods (mostly computer-assisted, on varied levels of sophistication and approximation finesse) to address
the spectral solution of  local   1D-3D  Schr\"{o}dinger operators in various areas of quantum physics  and quantum chemistry.    Special emphasis
is paid  there  to  low-lying  bound states. Here "low-lying" actually means that  numerically even few hundred of them are computable.

  Interestingly, with a notable exception of   relativistic  Hamiltonians where a numerical  methodology has been tailored to this specific
 case only,    no special attention has
  been paid to an obvious possibility to extend  these methods to   general nonlocal (and fractional in this number)
   energy operators,  which stem directly  from the L\'{e}vy-Khintchine formula.  The major  goal of  the present  paper  is  to   provide {\it  exemplary}
    numerically-assisted spectral   solutions  to   the latter  case,  while taking   fully  into  account  the inherent spatial nonlocality of the problem.

Efficient  2D and 3D  generalizations of computer routines  have been
worked out for local Hamiltonians. They rely on   higher  order factorizations of the semigroup operator \cite{Auer}-\cite{Chin}.
 An extension to nonlocal operators is here  possible  as well. A reliability  of the method
(including an issue of its sensitivity upon  integration  volume cut-offs)
  does not  significantly depend on the  particular  stability index $\mu $ or the
replacement of a  stable generator by the  relativistic one.

In the present paper we shall  focus on  1D   Cauchy-Schr\"{o}dinger  spectral problems  in two specific  confining regimes.
First we consider  an  analytically solvable  Cauchy oscillator problem,   \cite{SG,LM,R2,GS1},  which is
 viewed as  a test model for   an analysis of  possible intricacies/pitfalls  of  adopted   numerical procedures.   Next we shall pass
to the  Cauchy-Schr\"{o}dinger   well   problem. The well  is assumed to be   finite, but eventually  may become arbitrarily    deep. That will set a connection
with the infinite well problem,  considered so far in the  fractional QM  literature
 with rather limited success, c.f. \cite{Z}-\cite{K}). A   consistent   spectral  solution  of the  Cauchy   well  problem  is  our major task.

All computations will be carried out in  configuration space, thus deliberately  avoiding a
customary  usage of Fourier transforms  which definitely  blur the spatial nonlocality of   the problem. It is our aim
 to keep  under control   the balance between   the nonlocality impact  and  bounds upon the spatial
integration volume that are unavoidable in  numerical routines.  By  varying   spatial cutoffs we can  actually test  the
  reliability of the  computation   method,  e.g. the  convergence  towards  {\it would be  exact} eigenvalues and eigenfunctions. Here interpreted to arise
  as asymptotic  "Euclidean time"  limits.

  In addition  to  deducing  convergent approximations   for   lowest
 eigenvalues of confining fractional QM spectral problems,   we aim at  determining the  (convergent as well)  spatial   shapes of the   corresponding  eigenfunctions.
 That will  actually  resolve previously mentioned  inconsistencies (faulty or doubtful results)  in  the
  literature  devoted to fractional quantum mechanics,  see  in this connection   Refs. \cite{GS,YL}.

To this end,  we  adopt
  a computer-assisted route  to solve  the spectral problem for energy  operators   of the form  $H= T+V$ where $T $
  might be nonlocal while $V$ is  a locally defined external potential.  Its crucial ingredient is  an approximate form of
   the Trotter formula,  named the Strang splitting of the operator $\exp(-t H)$ with $0< t \ll 1$.
The Strang method has been originally devised for    local  Hamiltonians,  \cite{BBC}-\cite{Chin}.
    It is valid for short times and  not as a sole  operator identity, but  while   in action on suitable functions in the domain of $H$.

 The  main  advantage  of the Strang method  in the nonlocal context,  lies in an  easy  identification   of pitfalls   and  errors
   in the existing   theoretical discussions/disagreements   pertaining to spectral issues.
  More than that, correct answers to a number  of disputable points  can   be  given  with a  fairly
   high (numerical) finesse level, thus providing  a reliable  guidance  for  any  future   research
 on other fractional and relativistic  QM spectral problems.

The paper is structured as follows.
 In  Section 2 an outline is given of the  Strang splitting idea, widely used
in the context of spectral problems for  (local) Schr\"{o}dinger  operators.
Next we describe the  algorithm yielding an efficient simulation of an  initially given set of  trial  $L^2$ functions
towards  an asymptotic orthonormal set.  In Section 3 we extend the applicability range of the
 algorithm to nonlocal motion  generators which are the major concern of the present paper.

 In Section  3 we give a detailed study of the algorithm workings in comparison with known analytic results
 for the  1D Cauchy oscillator problem.    That includes an analysis  of an impact of spatial cutoffs (necessary for executing
  numerical routines) on obtained spectral data, while set against those obtained analytically for  the inherently
   nonlocal motion generator.

Section 4 contains a discussion of the finite  Cauchy  well spectral problem  and that of  a dependence of
resultant spectral data  on the well depth and the spatial range of integrations
 (cutoffs  which somewhat "tame"  a   nonlocality of the problem).  Links with the infinite  Cauchy
  well spectrum are established.
For the latter problem  we have decisively disproved   spectral solutions proposed so far in the literature.
They are wrong. This statement extends to all hitherto proposed spectra of fractional generators in the
infinite well (or "in the interval" as mathematicians use to say).
 A constructive part of our research is a fairly accurate solution for both  the   finite  and infinite Cauchy well spectral problem in case of lowest
 eigenstates and eigenvalues.

\section{Solution of the Schr\"{o}dinger eigenvalue problem by means of the Strang splitting method}

\subsection{The direct  semigroup approach}

 As far as the spectral solution for a self-adjoint   non-negative  operator
   $H$ is concerned, it is   the "imaginary time propagation"  i.e.
    the semigroup dynamics $\exp (-tH/\hbar )$ with  $t\geq 0$ that
particularly  matters, \cite{BBC,AK}. This, in view of  obvious domain and
convergence/regularization  properties which   are implicit in the
Euclidean/statistical (e.g. the partition function evaluation)
framework.   Subsequently we  scale away all dimensional constants to make further
arguments more  transparent (then e.g. $\hbar \equiv 1$).

Let  us consider the spectral problem for  $H$  of the form  $H = T + V$:
 \be H\,\psi_i(x) =
E_i\psi_i(x),\qquad i=1,2,\ldots, \ee
 where  $T$   is \it  not \rm
 necessarily a local differential operator (like the negative of the
Laplacian), but may be a nonlocal pseudo-differential operator  as
well.

For concreteness we mention that in below we shall mostly refer to $T$  as the  $1D$
 Cauchy operator which is nonlocally defined  as follows:
  \be
  T\,\psi(x) =
(-\Delta)^{1/2}\,\psi(x)=\frac{1}{\pi}\int\frac{\psi(x)-\psi(x+z)}{z^2}dz.
\ee
A technical subtlety of Eq. (2) is that  the integral is interpreted
in terms of its  Cauchy principal value.  Clearly,  $H$  needs  to
be considered as a self-adjoint operator  and its spectrum is expected to belong to
$R_+$,  \cite{SG,LM}.
 Then, the  one-parameter  family  $\{e^{-tH},\>t\geqslant 0\}$ constitutes
 a  (strongly continuous symmetric)  semigroup of interest.

 We shall confine further discussion to  simplest   cases   of confining
 potentials (the 1D harmonic potential $V(x)=x^2$ provides an example)  or confining boundary data
   (finite, eventually very deep well), such that
 the discrete  spectrum of $H$   is  strictly positive and   non-degenerate:
   $0<E_1 < E_2 < E_3 < \ldots $. The latter restriction may be lifted,
  since it is known how to handle degenerate spectral problems, \cite{BBC,AK}.

Let there be given   in
 $L^2(R)$  an orthonormal  basis   composed   of eigenfunctions $\psi_i(x)$  of  $H$, such that
 $[e^{-tH}\psi_i](x)=e^{-t E_i}\psi_i(x)$  for all $i\in N$.
Let  $\psi(x,0)=\psi(x)$ be a  pre-selected  {\it trial function} which is an   element of
$L^2(R)$.
 In the eigenbasis of $H$, for $\psi $ belonging to the domain of $H$, we have
$\psi =\sum\limits_{i}c_i\psi_i $, where  $
c_i=<\psi_i|\psi >$ and   $<\cdot|\cdot>$  denotes  the
$L^2(R)$  scalar product.

The evolution rule  reads     $\psi \rightarrow \psi (t)  = \exp(-H t)\, \psi (0)$  (alternatively $\partial _t \psi (x,t) =
-H \psi (x,t)$),  with $\psi(x,0)= \psi (x)$.   Accordingly, in the large time (albeit finite)
asymptotic (presuming that $c_1 \neq 0$)     we    have
  \be
\psi(x,t)=[e^{-tH}\psi ](x,0)=\sum\limits_{i} e^{-t E_i}c_i\psi_i(x)
\sim     e^{-t E_1}c_1\psi_1(x). \label{state}
\ee

If we do not a priori know   the   ground state  $\psi _1$  of $H$, it is the  $L^2(R)$-normalization of $\psi (t)$  for large times
 that   does the job, i.e.  yields   $\psi _1$.    The corresponding ground-state eigenvalue  can be  obtained by
computing   $<\psi _1|H\psi _1> = E_1$.

We note however that if  we   select $\psi (0)$   blindly (e.g.  plainly at   random), then it may happen that   $\psi (0)$  is
$L^2$-orthogonal to $\psi _1$ and the ground state surely will never  emerge in the  asymptotic   procedure  Eq. (\ref{state}). Instead, we
 would  arrive at   a certain  (lowest possible)
excited eigenstate of $H$.  Therefore  to   identify consecutive eigenvalues and eigenfunctions of
$H$,  a systematic (and optimal)  strategy  for a proper choice/guess of trial functions  $\psi (0)$
  appears  to be  vital.

\subsection{The  Strang method}

The Strang  splitting  method amounts to   approximating  a semigroup  operator,
 that yields the dynamics     $\exp (-tH) \psi $ of  a suitable  initial data vector
 $\psi $ for arbitrary $t>0$,  by a composition of  a large number of consecutive  small "time  shifts".
Its   core    lies in the  Trotter-type  splitting  of the semigroup  operator
$\exp (- H \Delta t)$, where $H= T+V$  and $\Delta t \ll 1$,  into   products of the form
\be
\mathcal{U}^{(p)}(\Delta t)\equiv \prod\limits_{i=1}^m e^{-a_i V \Delta t} e^{-b_i T \Delta t}.\nonumber
 \ee
Here  $\mathcal{U}^{(p)}(\Delta t)=e^{-H \Delta  t}+\mathcal{O}({\Delta t}^{p+1})$  is
 regarded as the $p$-th order approximation of $e^{-  H \Delta t} $, provided  $\Delta t$ is   sufficiently small,  $m$ in turn
 being  not too small and  the   coefficients $a_i, b_i$, that determine the approximation accuracy, need to
   obey   suitable  consistency conditions, see \cite{BBC}. For $p\leq 2$, positive
  $a_i, b_i$ are always admitted.

In the present paper we shall focus on  the  simplest, second order Strang   approximation, widely used \cite{Auer} in
physics and quantum chemistry contexts:
 \be \mathcal{U}^{(2)}(t)\equiv
e^{-\frac{t}{2}V}e^{-t\,T}e^{-\frac{t}{2}V}, \label{l1} \ee
where  there holds $\mathcal{U}^{(2)}(t)=e^{-tH}+\mathcal{O}(t^3)$.

Like in the standard  quantum mechanical  perturbation theory, the interpretation of  the $\mathcal{O}(t^3)$
 term as "sufficiently  small" remains somewhat obscure, unless
specified with reference to its action in the   domain of definition.
 \\

{\bf Remark 3:} As mentioned before, the approximation formula  (\ref{l1})  for the
semigroup operator $e^{-t H}$ is one   specific choice    among many
conceivable  others.   An example  of  the fourth rank approximation reads:
 \be
\mathcal{U}^{(4)}(t)\equiv
e^{-\frac{t}{6}V}e^{-\frac{t}{2}T}e^{-\frac{2t}{3}V-\frac{t^3}{72}[V,[T,V]]}e^{-\frac{t}{2}T}e^{-\frac{t}{6}V}.\nonumber
\ee A number of other  approximation formulas  can be found in Refs.
\cite{BBC,AK} together  with a discussion of their usefulness,
 see also \cite{A} for alternative considerations.\\

We note that an optimal  value of a "small" time shift unit  $\Delta t$, here by  denoted  $h$,
 appears to be model-dependent.   Subsequently, we shall   refer to   $h=0.001$,  which
 proves to be sufficient  for  third  and higher rank terms of the Taylor expansion of (\ref{l1})
  to  be considered negligible.  That in the context of the Cauchy oscillator and the
   Cauchy well  spectral  problems.

  A   preferably long  sequence     of consecutive small  $h$   "shifts"    of
    an initially given function $\psi(x,0) \rightarrow \psi (x,kh)$ with  $k= 1, 2, ...$,
    mimics  the  actual continuous evolution
    of $\psi (x,t)$ in  the time interval $[0, kh]$.
With the above assumptions,  the exact evolution operator $e^{-t H}$  is here
 replaced by an approximate expression, valid  in the regime of  sufficiently small times $t\sim h$:
  \be e^{-h H}\approx
e^{-\frac{h}{2}V}\left(1-h
T\right)e^{-\frac{h}{2}V}\equiv\mathcal{S}(h).\label{l2} \ee

 The induced approximation error
depends on the time step  $h$ value. If $h$ is small,  the error is small as well but
  the number of iterations towards  first  convergence symptoms is becoming large. Thus a proper balance between
  the two goals, e.g. the  accuracy level and the  optimal convergence performance, need to be established.
  (One more source of inaccuracies   is rooted in the nonlocality of involved operators and
  spatial cutoffs needed to evaluate integrals. This issue we shall discuss later.)

  An outline of the  algorithm that is appropriate for a numerical implementation
   and ultimately  is capable of   generating   approximate  eigenvalues and eigenfunctions
    of $H$, reads as follows:

  (i) We  choose a finite number  $n$
  of    trial  state  vectors (preferably linearly independent) $\{ \Phi_i^{(0)}, \, 1\leq i\leq n\} $, where
     $n$ is correlated with an ultimate number
   of eigenvectors of $H$ to be obtained in the numerical procedure;
      at the moment we disregard an issue of their optimal (purpose-dependent) choice.

   (ii)  For all trial functions   the time  evolution beginning    at $t=0$ and
terminating at  $t= h$, for all $1\leq i \leq n$  is  mimicked by the time shift  operator
$S(h)$
 \be \Psi_i^{(1)}(x)=S(h)\Phi_i^{(0)}(x). \ee

 (iii) The obtained set of  linearly independent  vectors
$\{\Psi_i^{(1)}\}$ should be made orthogonal  (we shall use the  familiar
Gram-Schmidt  procedure, although there are many others,  \cite{AK})  and normalized. The
 outcome  constitutes a {\it new} set   of trial states
 $\{ \Phi_i^{(1)}, i=1,2,\ldots, n\}$.

(iv) Steps (ii)  and (iii) are next repeated  consecutively, giving rise to a temporally ordered sequence
of    $n$-element   orthonormal   sets  $\{ \Phi_i^{(k)}(x),
i=1,2,\ldots, n\}$ and the resultant  set of linearly independent vectors
\be
\Psi_i^{(k+1)}(x)=S(h)\Phi_i^{(k)}(x),\qquad i=1,2,\ldots,n,  \nonumber
\ee
 at time $t_{k+1}= (k+1) \cdot h$.  We main abstain from its orthonormalization  and stop the iteration procedure, if
 definite symptoms of convergence  are detected.
 A discussion of   operational convergence criterions can be found e.g.  in Ref. \cite{Chin}.

(v) The   temporally ordered  sequence of
$\Phi_i^{(k)}(x)$,  $k\geq 1$
  for  sufficiently  large  $k$  is   expected to converge to an eigenvector  of $S(h)$, according to:
\be S(h)\Phi_i^{(k)}(x)=
e^{- h E_i^{(k)}}\Phi_i^{(k)}(x) \sim   e^{-  h E_i}\psi_i(x),  \label{E} \ee
 where actually
$\psi_i$ actually   stands for an eigenvector of  $H$  corresponding to the
eigenvalue $E$.
Here:
\be
 E_i^{(k)}(h) = -\frac{1}{h}\ln(\mathcal{E}_i^{k}(h)),\label{l3}
 \ee
where
\be
\mathcal{E}_i^{k}(h)=<\Phi_i^{(k)}|\Psi_i^{(k+1)}>=<\Phi_i^{(k)}|S(h)\Phi_i^{(k)}>,\nonumber
\ee
is  an expectation value of   $S(h)$ in the $i$-th state $\Phi_i^{(k)}$.

 It is   the evaluation of  $\Phi_i^{(k)}(x)$ and    $E_i^{(k)}(h)$ that is amenable to standard
  computing routines  and yields approximate eigenfunctions and   eigenvalues of $H$. The degree of
 approximation   accuracy is   set by the terminal  time  value $t_k=kh$,  at which
 the symptoms of convergence are  detected and  the iteration (i)-(v) is thence  stopped.

 In passing let us mention one  more  obvious (in addition to  the second order Strang splitting choice, instead of  higher order formulas)
   source of the approximation inaccuracy in the computer-assisted
 evaluation of eigenfunctions and eigenvalues.  Namely, in the Strang splitting formula  (5), the formal Taylor series for
  $\exp(-tT)= 1-tT + (t^2/2!)T^2 -...$   have been  cut after the   linear
 in $t$  term.

\subsection{Accounting for spatial cutoffs}

As mentioned before,  an important  source of inaccuracies of numerical procedures    is rooted in the
 nonlocality of involved operators and   in  spatial cutoffs needed to evaluate the integrals.
 Till now we have  introduced  $h=0,001$  as a  partition  unit  for any time interval in question.
 However,  to  execute any numerical  routine pertaining to  nonlocal operators of the   Cauchy  form  (2),
  we need to set an upper bound for the integration interval  and  additionally
  select an appropriate  spatial  partition unit.

In $1D$, from the start we  need to choose  $x\in[-a,a]$, $a>0$.
 How wide the spatial interval should be  to yield  reliable simulation outcomes,
 is a matter of a  numerical   experimentation.  Subsequently, we shall analyze how
 sensitive  the  simulation outcomes  are upon  a concrete   choice of  $a$.

For   computing purposes, the finite  spatial interval   must be   partitioned  into a large number
$N$  of  small subintervals.  The  partition finesse is crucial for
the fidelity of integrations. Once $a$ is fixed, by taking large $N$,  we considerably   increase the
simulation time. Therefore  we need to  set a   balance between the overall
computer  time cost and integration reliability.

In the present paper, the   spatial partition unit is set to be  $\Delta x=0.001$.  Consequently, for $a=1$, the
interval $[-1,1]$  is being  partitioned into $2 \cdot 10^3$ subintervals.

In view of the a priori declared  $[-a,a]$  integration  boundary limits,  irrespective of the
initial data choice  $\{ \Phi_i^{(0)} \in L^2(R)\}$,
 the simulation  outcome     is automatically placed in   $L^2([-a,a])$.

(We point out that what we deal with "behind the stage", because of the semigroup dynamics involved,  are
 heavy-tailed jump type processes. Their long jumps have  a considerable probability to occur, hence
  taming them  by spatial  cutoffs needs to be under scrutiny and subject to control.)

For  the Cauchy  oscillator whose eigenfunctions extend over the whole real line,  we
effectively get an approximation of  true eigenfunctions by functions with a support  restricted to  $[-a,a]$.
 Clearly, the value of  $a$ cannot be too small and  for the present purpose  the minimal value of $a=50$ has been
 found  to be  a reliable choice.
This point must be  continually  kept in mind.

\section{Cauchy oscillator: Strang method versus exact  spectral solution}

To test a predictive power of the just  outlined  computer-assisted
  method of solution of the Schr\"{o}dinger-type  spectral problem, while extended to a non-local operator $H$, we
shall take advantage of the existence in $L^2(R)$ of a complete analytic solution of the
Cauchy oscillator problem, \cite{GS,LM}.

 Remembering that  the   algorithm  (i)-(v) of Section II.B, even if started in $L^2(R)$  will necessarily place all our discussion
 in  $L^2([-a,a])$, we need to presume that
 $a$ is  "sufficiently large". We shall   subsequently   describe   how the simulation outcomes  for the Cauchy oscillator   spectral
solution     depend on the specific  choice of $a\geq 50$.\\

{\bf Remark 4:}  We recall  that  one  needs to be aware of the  relevance of long jumps
  (that in view of the heavy-tailed  L\'{e}vy  noise distribution function)
 in  the  jump-type process  which  underlies the semigroup dynamics.
    For an unrestricted (free, with    no boundary restrictions or external potentials in action) an impact of both above
    cut-offs has been investigated in Ref. \cite{SM}. Instead of the L\'{e}vy jump-type process, there   appears a standard jump process.
    Since L\'{e}vy measures are still involved, there is some preference of long jumps (heavy-tails issue).
    For all standard jump processes  a convergence  to a Gaussian (law of large numbers) is known to arise. In case of truncated L\'{e}vy processes
   an ultraslow convergence to a Gaussian has been reported   to occur \cite{SM}.
 This observation  does  not    extend  to  confined L\'{e}vy flights, even in their truncated version. They asymptotically set down
 at  heavy-tailed   pdfs   \cite{SG} and \cite{GS1}.

 For our purposes the natural  $L^2(R)$  choice   of initially given trial state vectors
 is that of Hermite functions
 \be \Phi_{i+1}^{(0)}(x)=\frac{1}{\sqrt{2^i
i!\sqrt{\pi}}}H_i(x)e^{-x^2/2},\qquad i=0,1,\ldots \ee
 where $H_i(x)$ are Hermite polynomials, defined by the Rodrigues formula
\be H_i(x) = (-1)^i e^{x^2}\frac{d^i}{dx^i}e^{-x^2},\qquad
i=0,1,\ldots\nonumber \ee
We recall that $H_0(x) = 1, H_1(x) = 2x,
H_2(x) = 4x^2-2, H_3(x) = 8x^3-12x$ and so on.

The functions (9) form a standard (quantum  harmonic oscillator)
basis in  $L^2(R)$. They loose this property after the first time-shift operation (6), being  mapped
into linearly independent functions with support restricted to  $[-a,a]$.
The resultant $L^2([-a,a])$    functions
 $\Psi_i^{(1)}(x)$  preserve  a  track of the
number of nodes  and  the  related  evenness/oddness    properties.

  At this point it is necessary to mention that the
  Courant-Hilbert nodal line theorem   has never  been extended to  operators which are nonlocal, \cite{BK}).
 As well, no its analog is known   in the current  context.
Nonetheless, our  simulation routines will reproduce  the standard nodal picture for approximate eigenvectors,
in consistency with previously established  analytic  properties of the Cauchy oscillator eigenfunctions, \cite{SG,LM}.

 \begin{figure}[h]
 \begin{center}
 \centering
 \includegraphics[width=70mm,height=70mm]{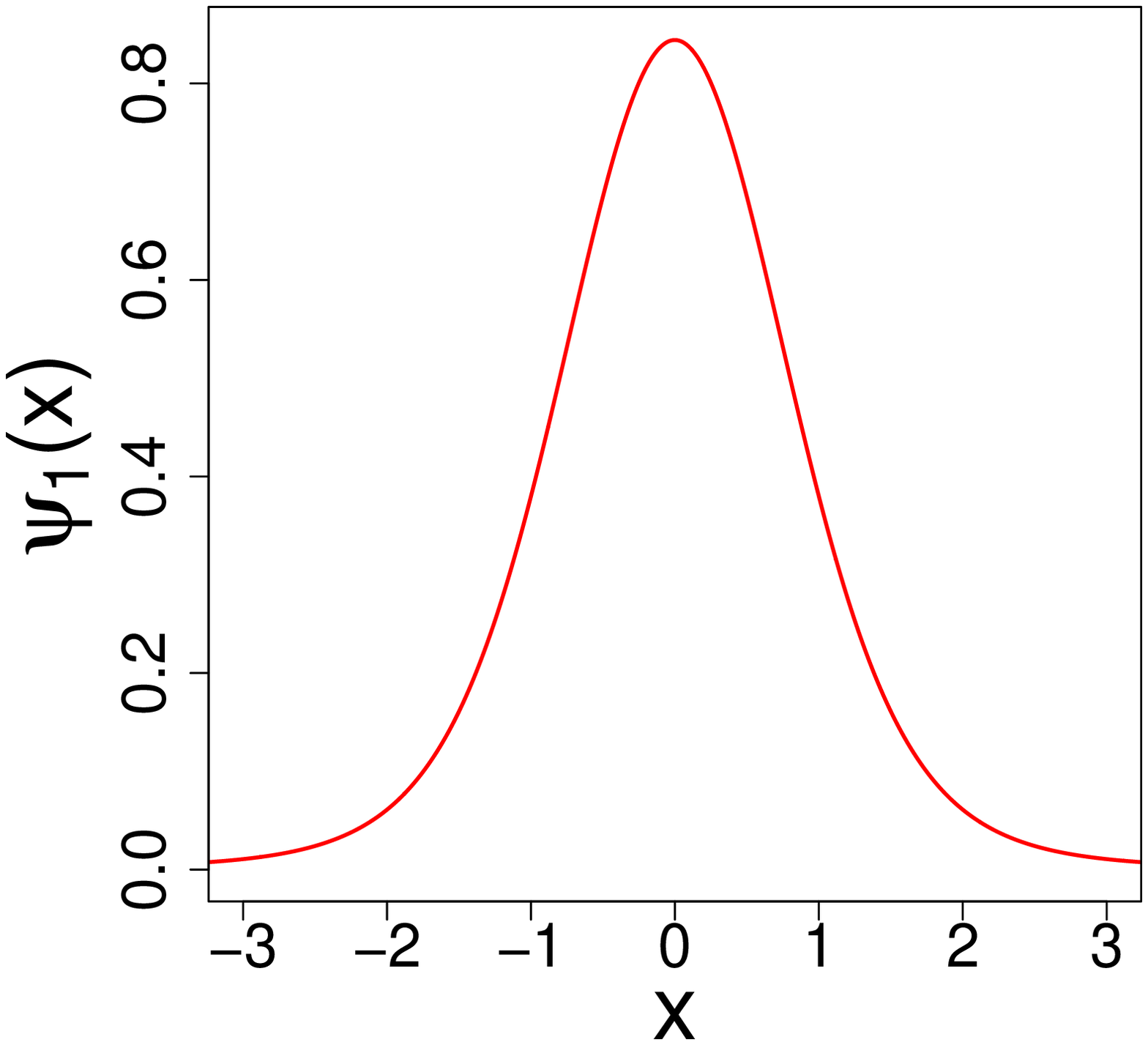}
 \includegraphics[width=70mm,height=70mm]{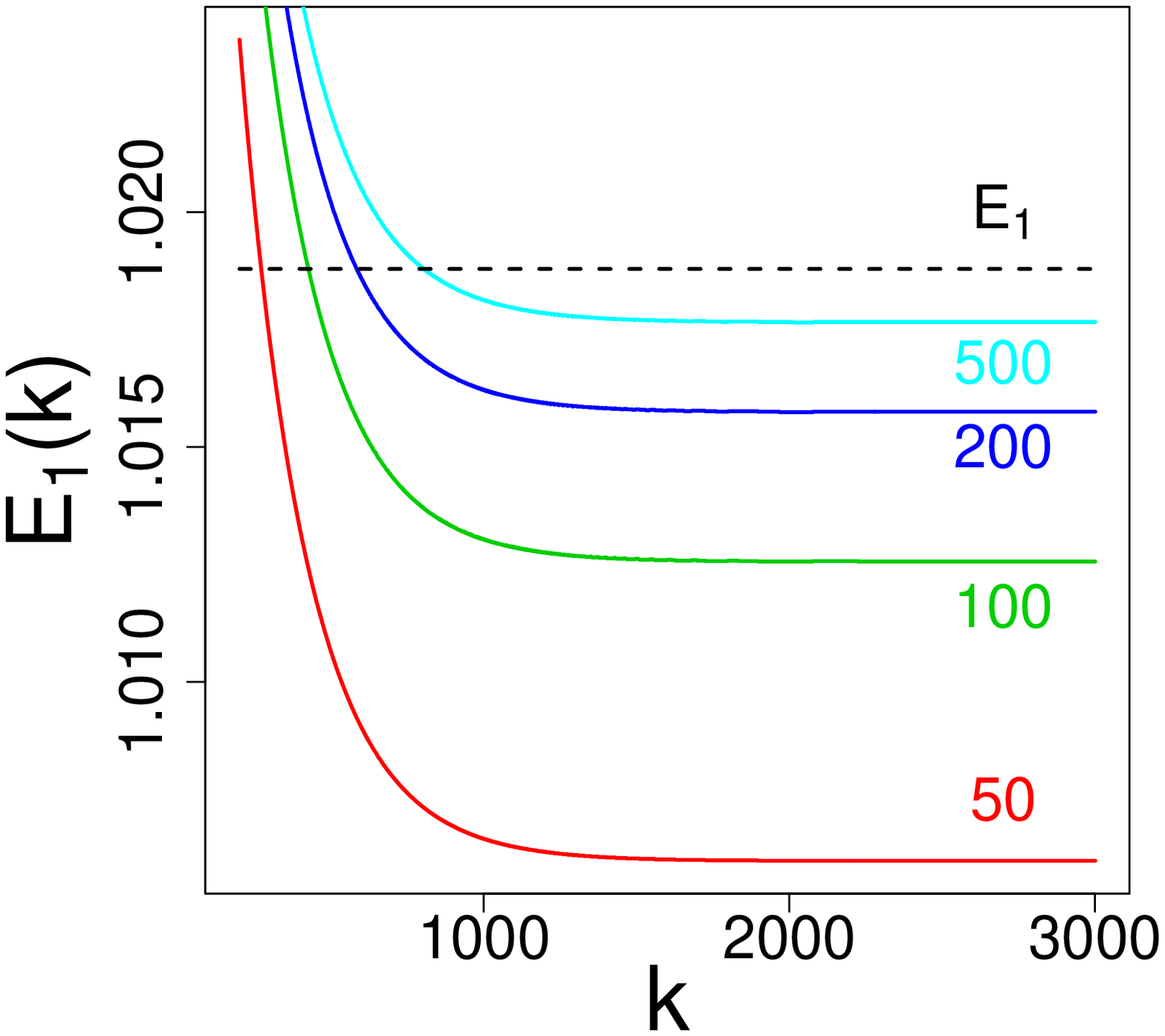}\\
 \caption{Cauchy oscillator ground state (left panel) and the   ($k$)-time  evolution of  $ E_1^{(k)}(h) = -\frac{1}{h}\ln(\mathcal{E}_1^{k}(h))$,
  (\ref{l3}),  for  $a=50, 100, 200, 500$.
  The dotted line indicates the ground state eigenvalue reported  in  Refs. \cite{LM,SG}.}
 \end{center}
 \end{figure}

\subsection{Cauchy oscillator ground state.}

As the initial  trial state vector we take the Hermite function   $\Phi _1^{(0)}(x)$, (9), which is  subsequently   evolved up to $k= 3000$, with $h= 10^{-3}$.
 In Fig. $1$, where the Cauchy oscillator ground state is  reproduced,   we  have  intentionally  abstained from   depicting   the corresponding
  $\Phi_1^{(k)}(x)$. The reason is that   within the  graphical reproduction accuracy limits,  in the adopted   scale and  with the
   spatial boundary set  by   $a\geq 50$,
  the  computed curve  cannot be   practically  distinguished from the Cauchy oscillator ground state  $\psi _1(x)$, \cite{SG}.   Note the
  visually obvious  irrelevance of the  boundaries: in Fig. 1  an interval $[-3,3]$ has been displayed, while  $[-50, 50]$ is
  actually  considered  in simulations.

   For completeness we provide  an analytic form of the Cauchy oscillator ground state (we recall a dimensionless form of this expression):
  \be
\psi_1(x)=\frac{A_0}{\pi}\int\limits_{-E_1}^\infty Ai(t)\cos
x(t+ E_1)dt,\label{l4} \ee
 Here, $E_1=1.01879297$,  while  $Ai(t)$  denotes the  Airy  function, $A_0$  being a normalization factor.

The   $k=3000$  approximation   $E_1^{(k)}(0.001)$   of a "true" eigenvalue  $E_1$ is depicted in Fig. $1$    for
 values  $a=50, 100, 200, 500$   of the boundary parameter. In this case,  scales have  been  modified  to
  better  visualize   the  impact of  $a$  (alternatively, a contribution of longer jumps  in the semigroup-driven stochastic process,
  as encoded in the involved L\'{e}vy measure)   upon the   approximation finesse  of the  lowest Cauchy oscillator eigenvalue  $E_1$.

A useful hint towards the  large  $a$ behavior of the detuning   (deviation)  $\Delta E_1 = E_1 - E_1^{(k)}$
 of the approximate energy value  $E_1^{(k)}$ from  the analytically  derived  value $E_1$   comes  from  Fig. 2, where a double  logarithmic
 (decimal logarithm is used on both axis) scale is employed.
   Clearly,  with the growth of $a$, the detuning value  $\Delta E_1$    decreases.

 Since  the Cauchy oscillator eigenfunctions  $\psi _n(x), n\geq 1$   obey an  asymptotic  estimate, \cite{KK,LM}:
\be
|\psi_n(x)|\leqslant \frac{C_n}{x^4},\qquad |x|>1, \label{asymptote}
\ee
 in Fig. 2 we have also displayed  the  graph of   $x^4\,\Phi_1^{3000}(x)$ for $a=500$.
 The   $x$-dependence    in Fig. 2  has been restricted to an interval $[-100,100]$, since beyond this interval
the  deviation of the graph from zero  quickly becomes {\it fapp}-negligible ({\it fapp} means
  "for all practical purposes").

\begin{figure}[h]
\begin{center}
\centering
\includegraphics[width=70mm,height=70mm]{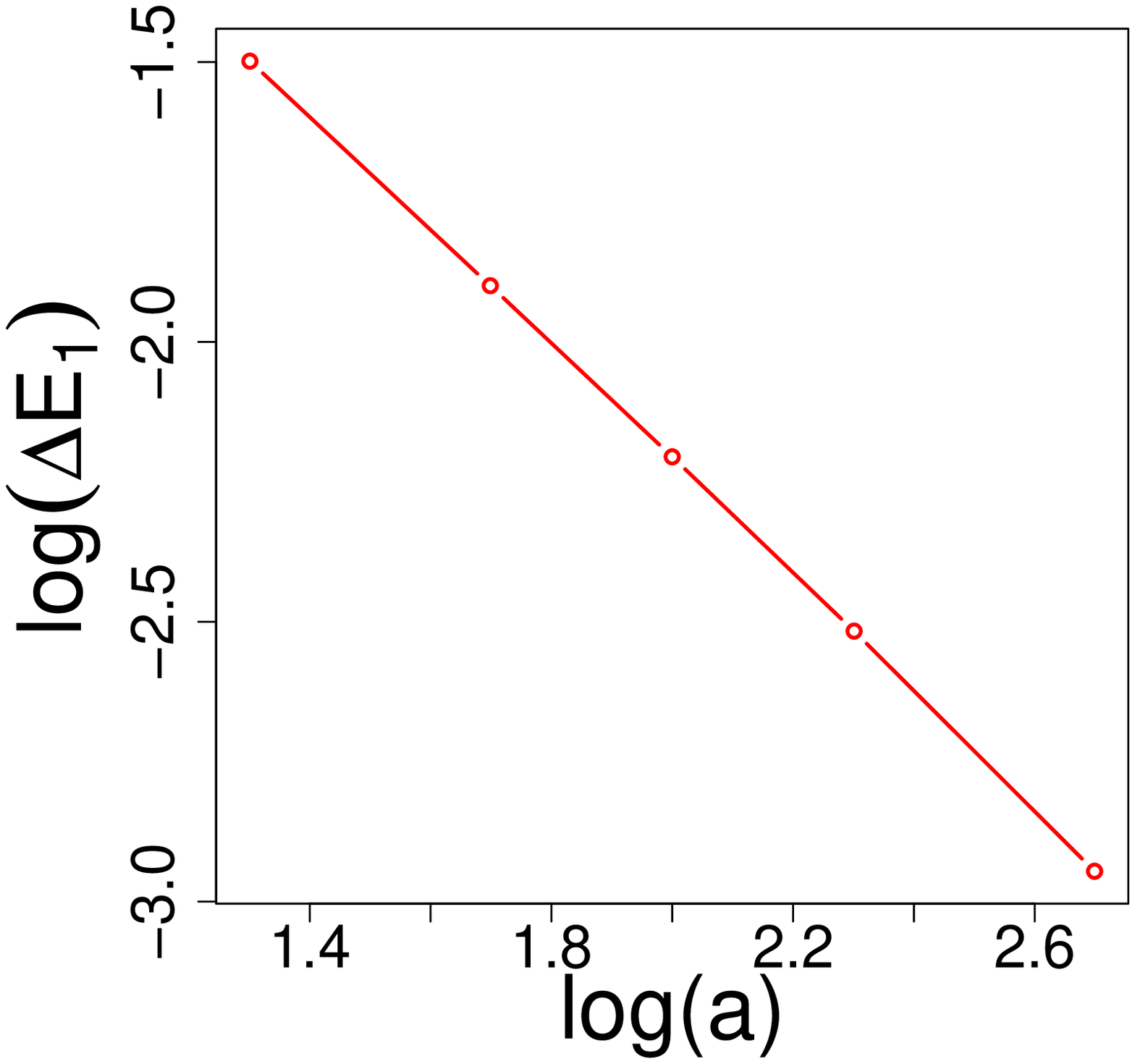}
\includegraphics[width=70mm,height=70mm]{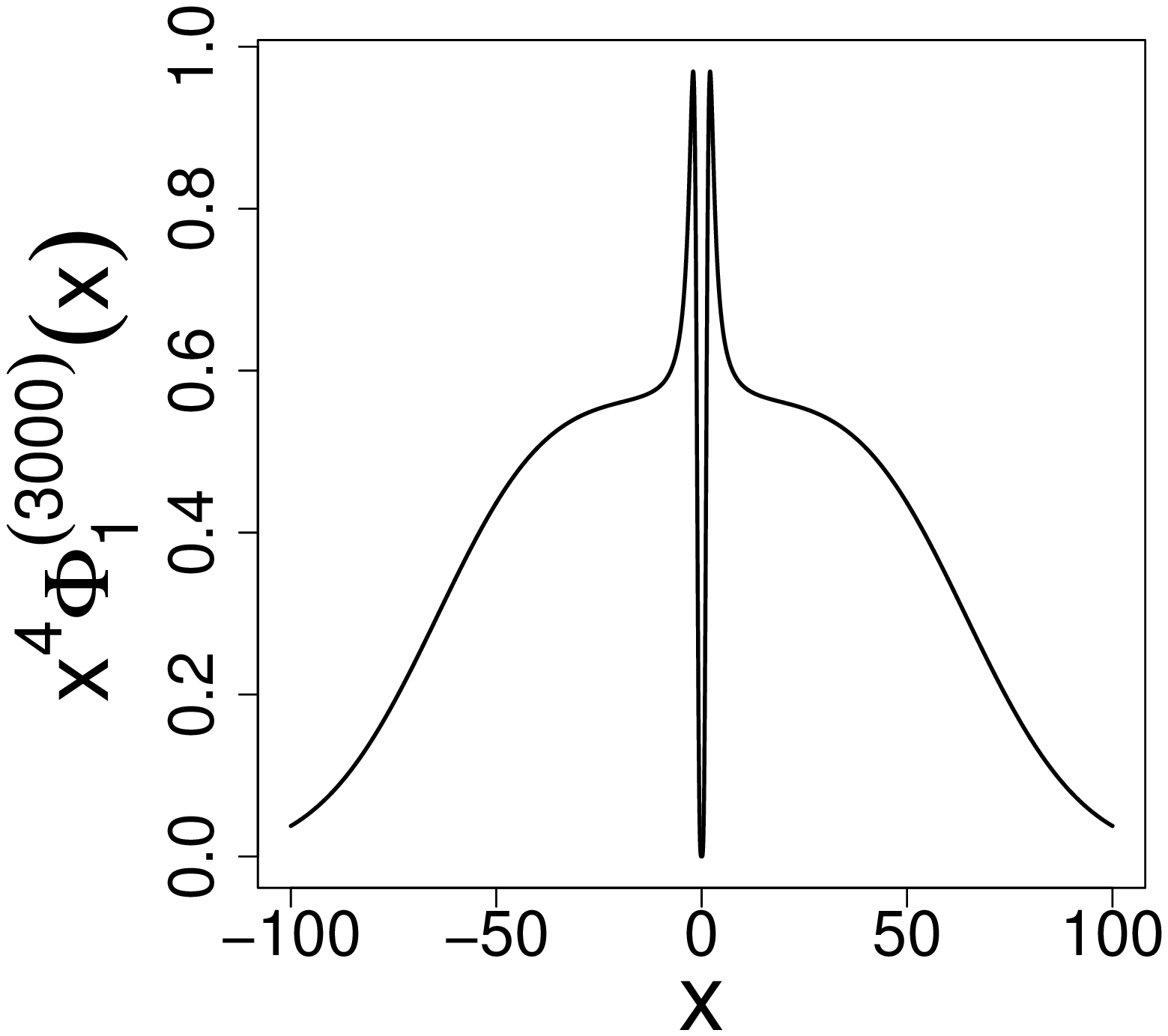}\\
\caption{ Detuning $\Delta E_1$-dependence upon $a$, in the (decimal) logarithmic scales  and the behavior of  $x^4 \Phi_1^{(3000)}(x)$ for $a=500$.}
\end{center}
\end{figure}

 \subsection{Lowest excited states}

\subsubsection{$n=2$}

From a formal point of view, while strictly complying with the point (i) of the algorithm (c.f. Section II.B), to recover simultaneously the ground state
and the first excited state, we  need to take two Hermite functions $\Phi _1^{(0)}, \Phi _2^{(0)}$  as our trial states. However, there are more optimal
choices as well  and, given $\psi _1$,   we can take a single Hermite function  $\Phi _2^{(0)}$  (having one node) as a trial state and
$(k)$-evolve it towards the sought for $\psi _2$.

In Fig. 3   the computed   state  function  $\Phi_2^{(k)}(x)$  is  depicted for   $k=3000$.  This outcome is practically ({\it fapp})
$a$-independent.
As in case of the ground state,   spatial  boundaries in the Figure  are set at $[-3,3]$, while  computations  have been performed for $a=50$.

  An identification of
$\Phi_2^{(k)}(x)$  as a valid   approximation of a true first excited state  $\psi _2(x)$ is supported by two arguments:  (i) an analytic form of
this  excited state  is available,  and its graph cannot be visually distinguished form the depicted approximate graph,
 within scales adopted, (ii)  the $k$-evolution of $E_2^{(k)}$  for different choices  of $a$,
displayed against  the analytically supported value   $E_2\approx 2,3381$   (dotted line), c.f.  \cite{LM, SG}.
\begin{figure}[h]
\begin{center}
\centering
\includegraphics[width=70mm,height=70mm]{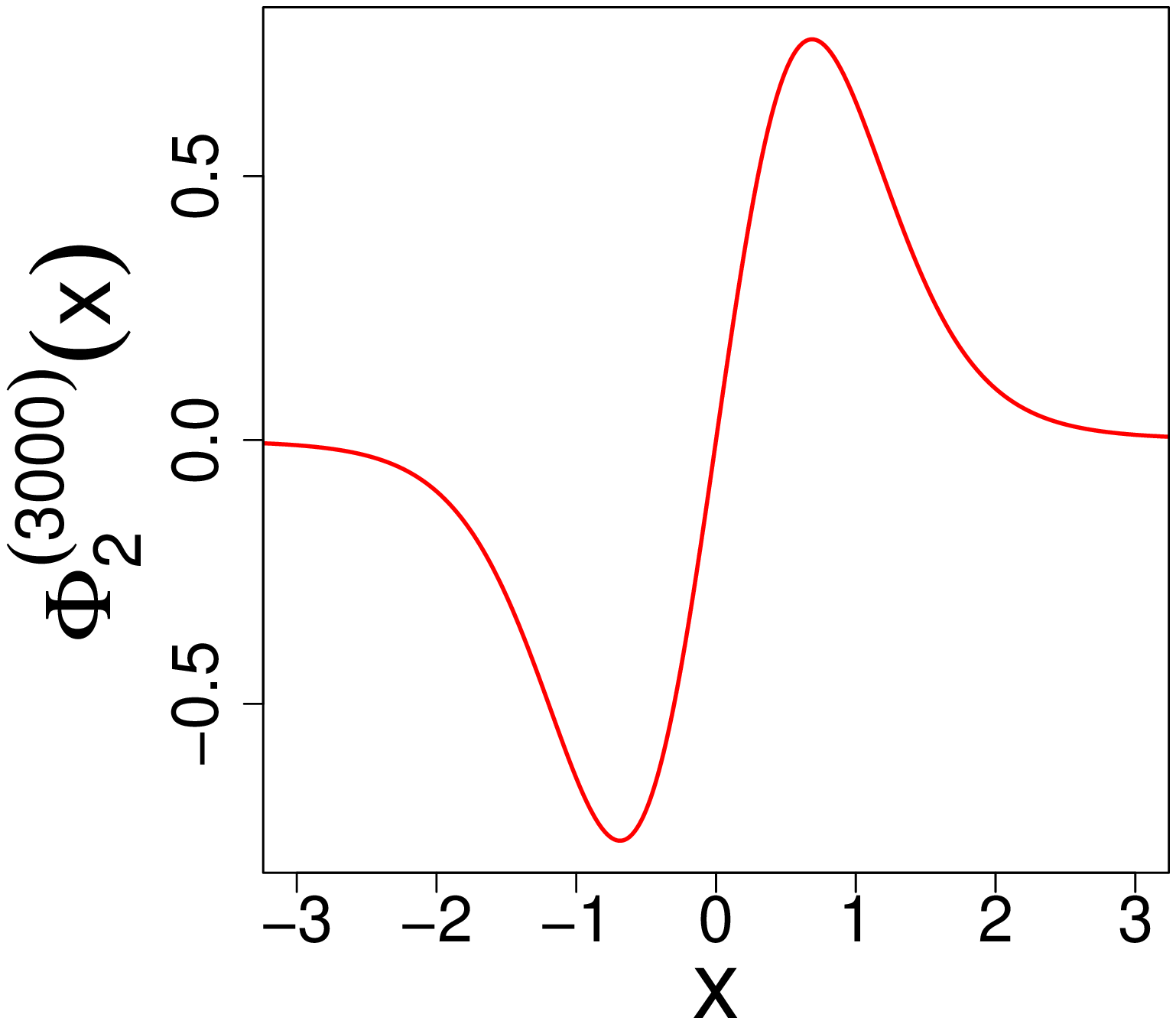}
\includegraphics[width=70mm,height=70mm]{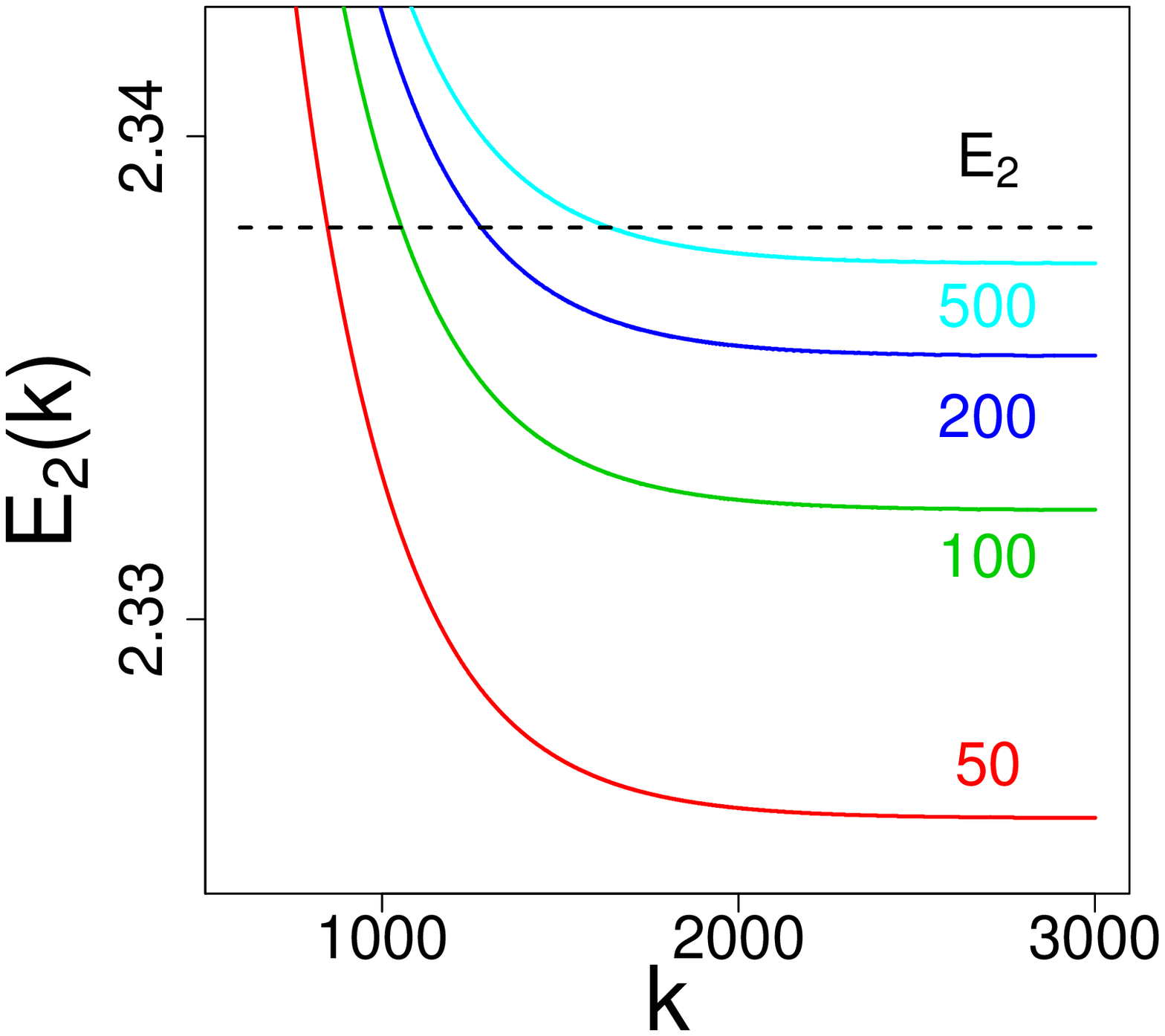}\\
\caption{First excited state  $\psi _2$ of the Cauchy oscillator  and  the  ($k$)-time  evolution of  $ E_2^{(k)}(h)$  (\ref{l3})
 for  $a=50, 100, 200, 500$.
The dotted line  indicates the  first excited  eigenvalue reported  in  Refs. \cite{LM,SG}.}
\end{center}
\end{figure}
In Fig. 4 we  have verified  the validity of an  asymptotic estimate (\ref{asymptote})  and the detuning
$\Delta E_2= E_2 - E_2^{(k)}$ - dependence on $a$
 in a double logarithmic scale.

Let us stress that the nodal properties of the computed eigenfunction are a consequence of the  oddness  of the chosen trial function and {\it  do  not}
trivially  rely on the initial (trial function)  number of nodes. For example,  if instead of the previous odd function  $\Phi_2^{(0)}(x)$
 we would have considered its spatially translated  $x\rightarrow x-1$ version
\be
\frac{1}{\sqrt{2\sqrt{\pi}}}\,2(x-1)\,e^{-(x-1)^2/2},\nonumber
\ee
as a trial function, then a simulation  outcome would  converge to the $E_1$ -  ground state  $\psi _1$
 and not (as possibly anticipated) to the excited $E_2$ - state $\psi _2$  .
\begin{figure}[h]
\begin{center}
\centering
\includegraphics[width=70mm,height=70mm]{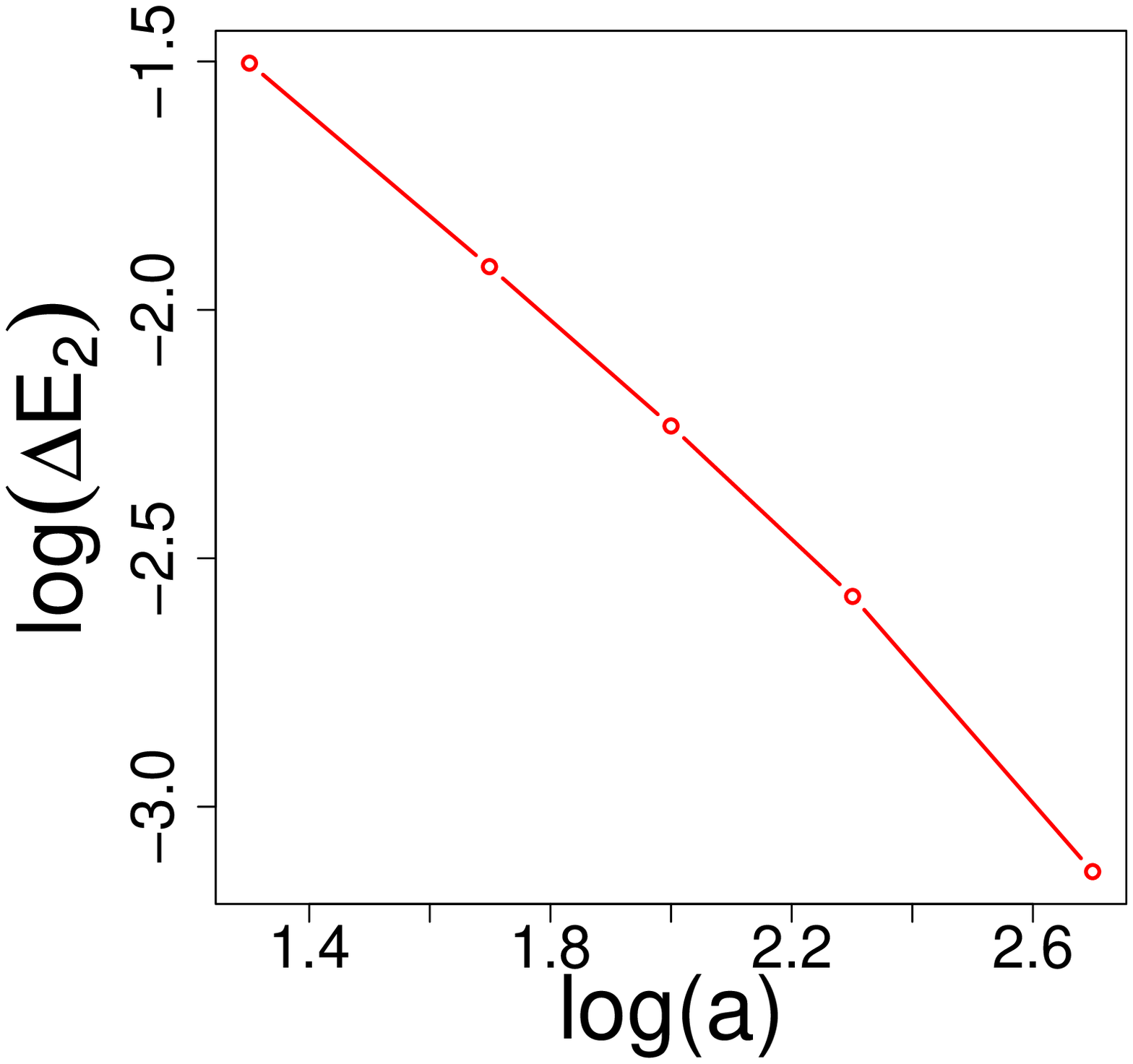}
\includegraphics[width=70mm,height=70mm]{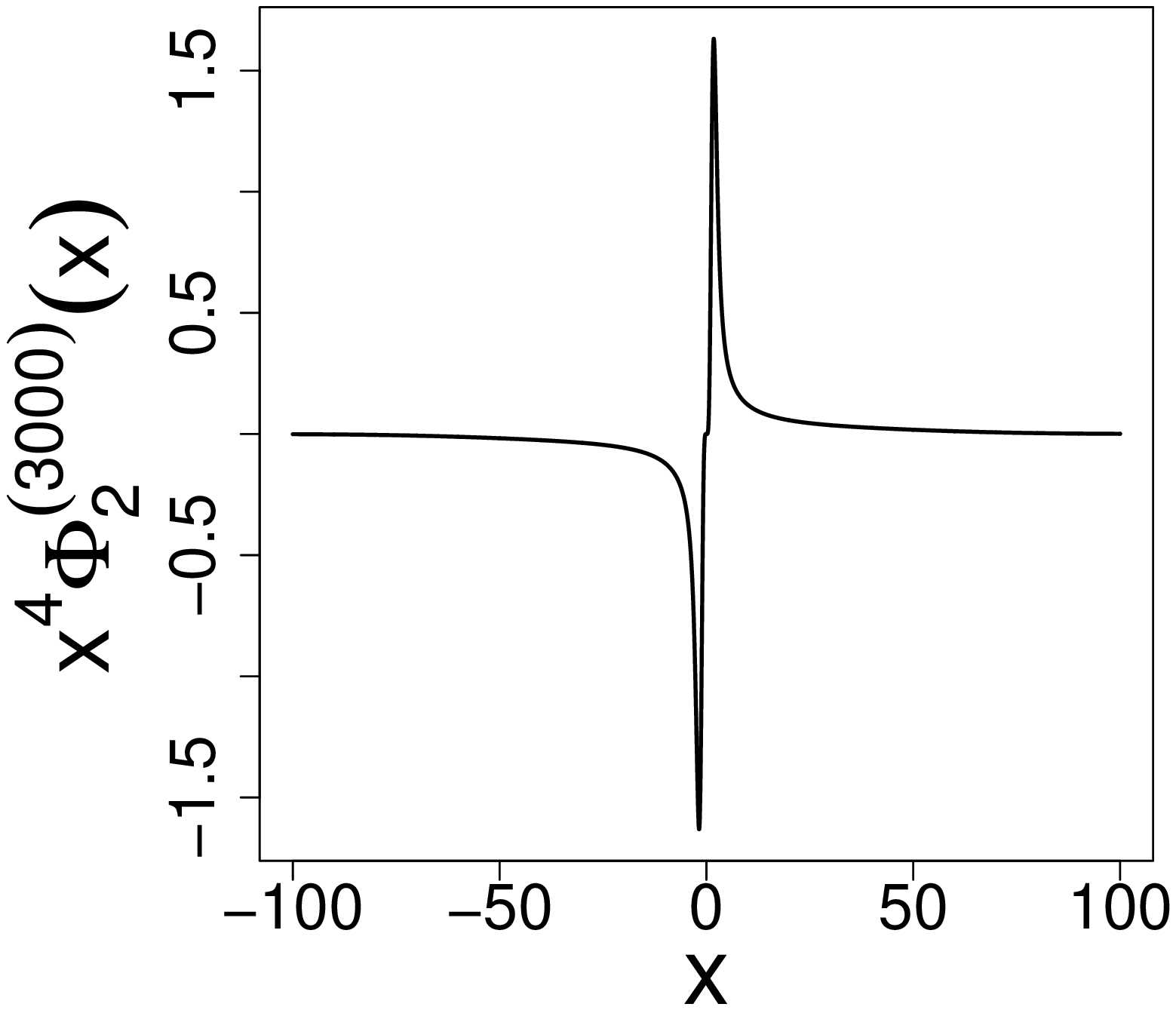}\\
\caption{Detuning $\Delta E_2$-dependence upon $a$, in the (decimal) logarithmic scales  and the behavior of  $x^4 \Phi_2^{(3000)}(x)$ for $a=500$.}
\end{center}
\end{figure}

In tune with the previous observation, let us notice that the  $i=2$  Hermite  function is even and has two nodes.
  If taken as a trial function for  our  algorithm,
it converges to the Cauchy oscillator ground state  and not to any excited state.

\subsubsection{$n \geq  3$}

To  compute $n \geq 3$  eigenfunctions we must  begin from at least two linearly independent  even  trial functions  and follow  iteratively
 all steps  (i)-(v)  of the algorithm  outlined in Section II.B.
If we choose  as  trial  functions   (in the notation of Eq. (9)) a two element set comprising
$\Phi_1^{(0)}(x)$  and  $\Phi_3^{(0)}(x)$, then  the algorithm
produces a sequence of state vector pairs which converges to that composed of  the  ground state $\psi _1$ and  the  excited state $\psi _3$,
  the latter  corresponding to $E_3 = 3.24819$.

\begin{figure}[h]
\begin{center}
\centering
\includegraphics[width=70mm,height=70mm]{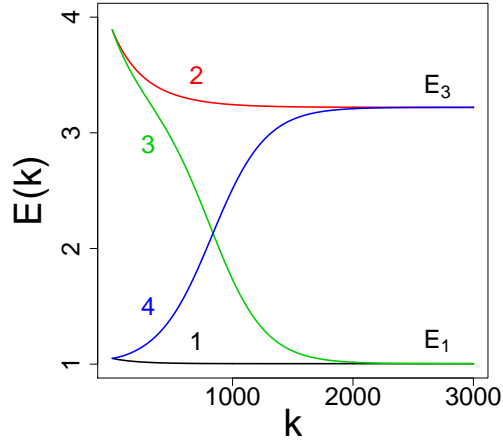}
\caption{Two ordering  choices in the Gram-Schmidt procedure. We depict the  ($k$)-time evolution of  $E_i^{(k)}$, (\ref{l3}),
  for initially chosen trial functions  $\Phi_1^{(0)}(x)$  and   $\Phi_3^{(0)}(x)$.  For   $E_1^{(k)}$  the  (1,3) G-S order induces the  black (1) curve, while  blue (4) for  (3,1) G-S order.
  For $E_3^{(k)}$  the (1,3) order yields the  red (2)  curve  while (3,1)  implies the   green (3) one.}
\end{center}
\end{figure}
\begin{figure}[h]
\begin{center}
\centering
\includegraphics[width=50mm,height=50mm]{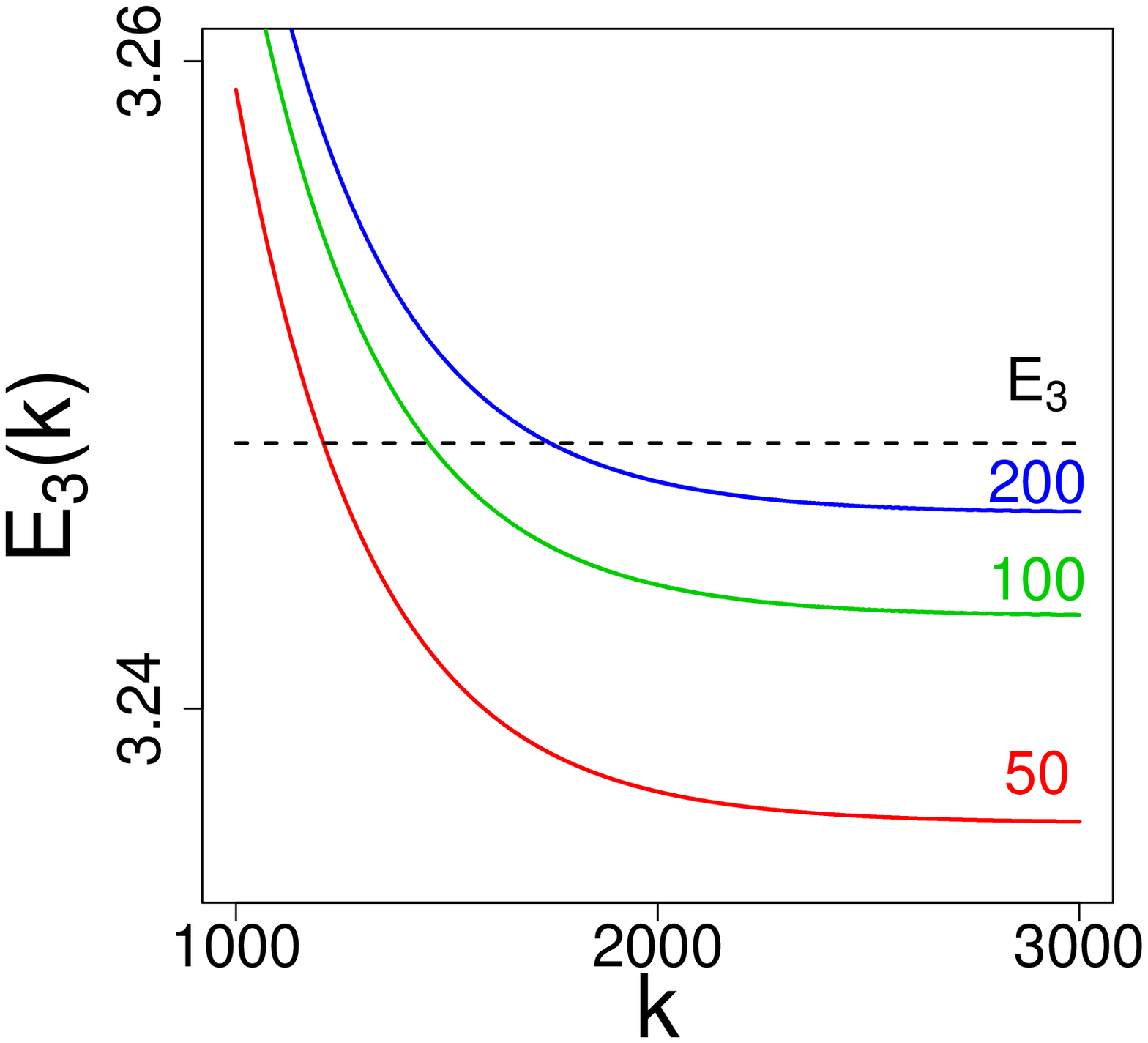}
\includegraphics[width=50mm,height=50mm]{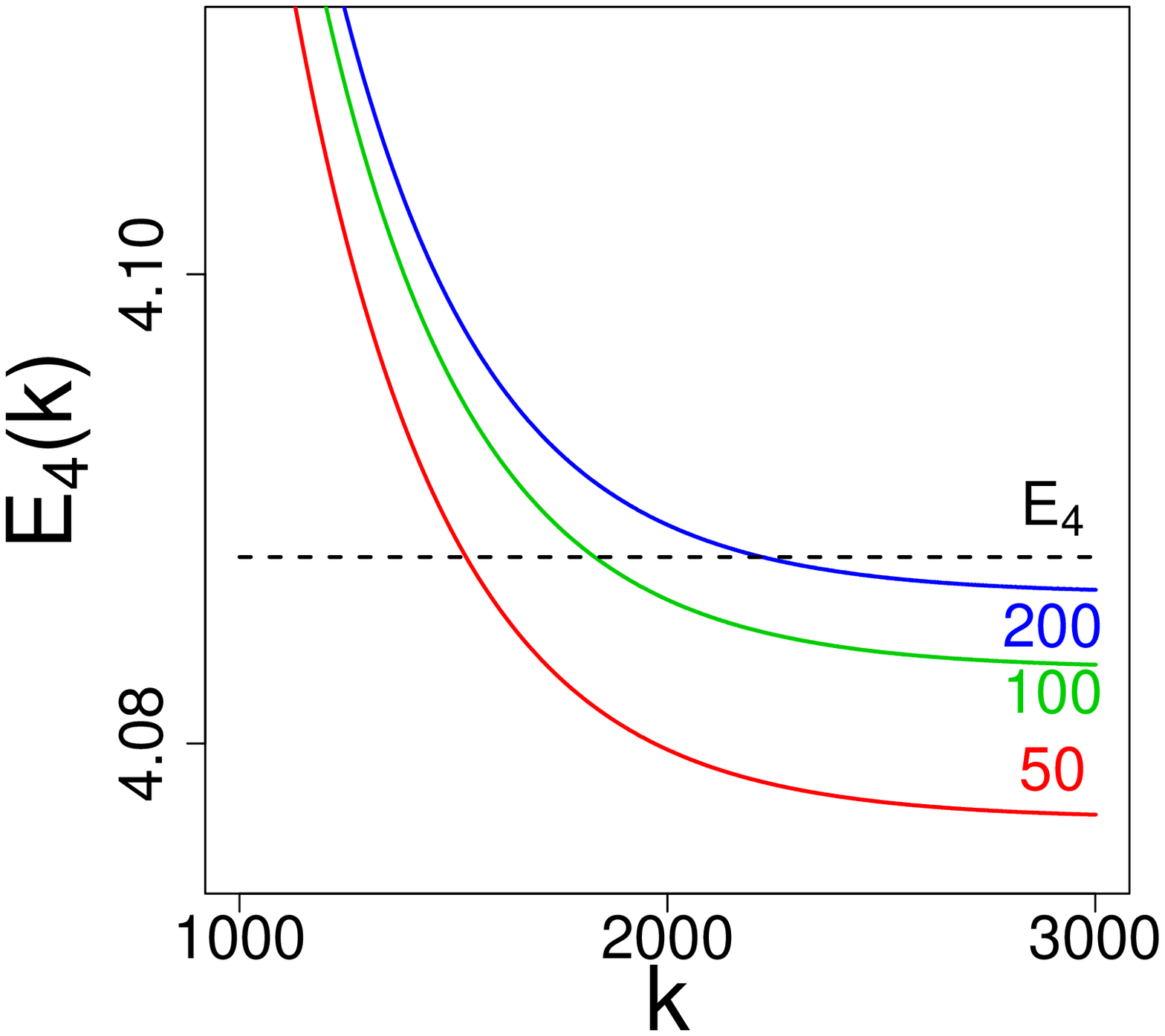}
\includegraphics[width=50mm,height=50mm]{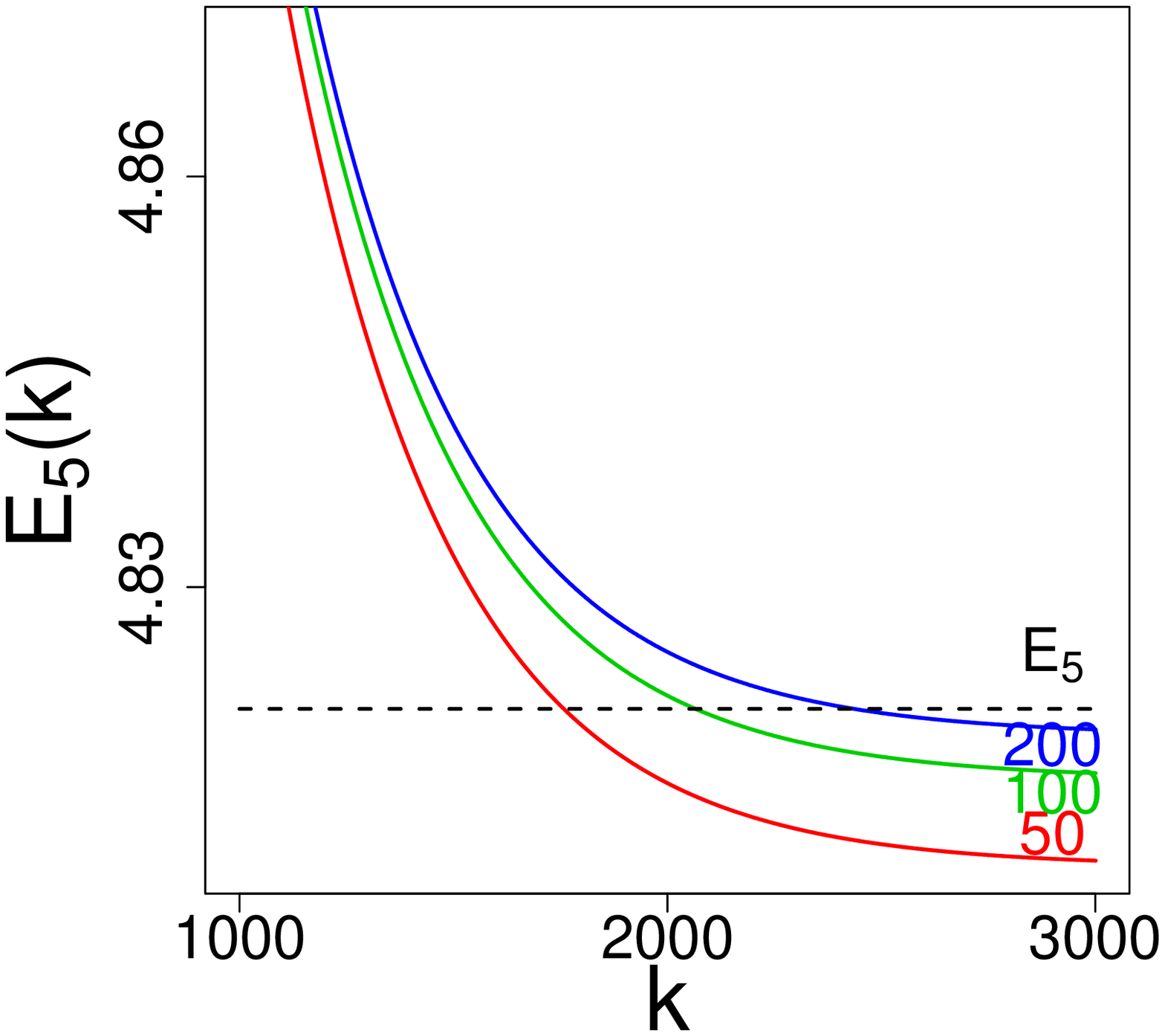}
\includegraphics[width=50mm,height=50mm]{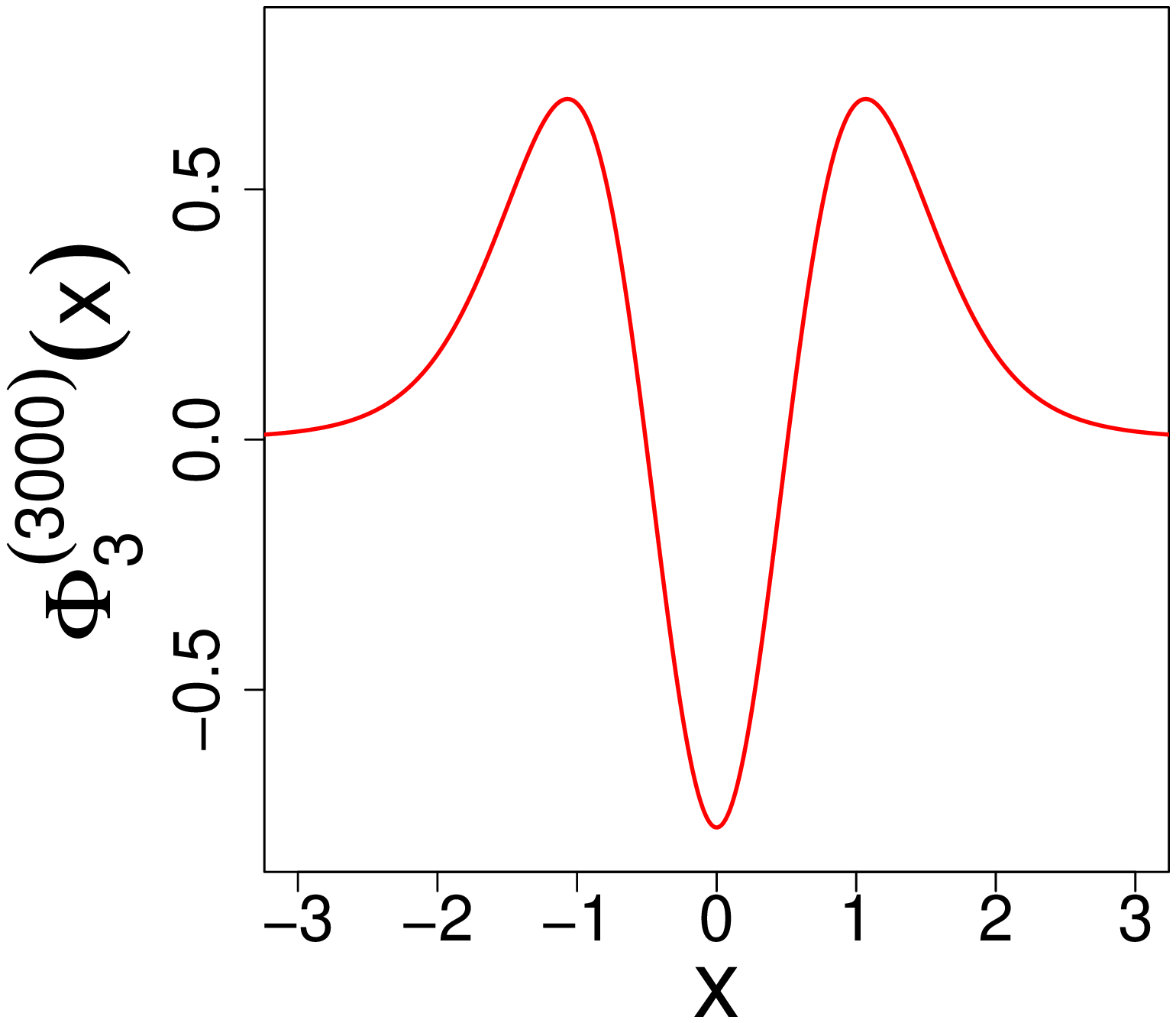}
\includegraphics[width=50mm,height=50mm]{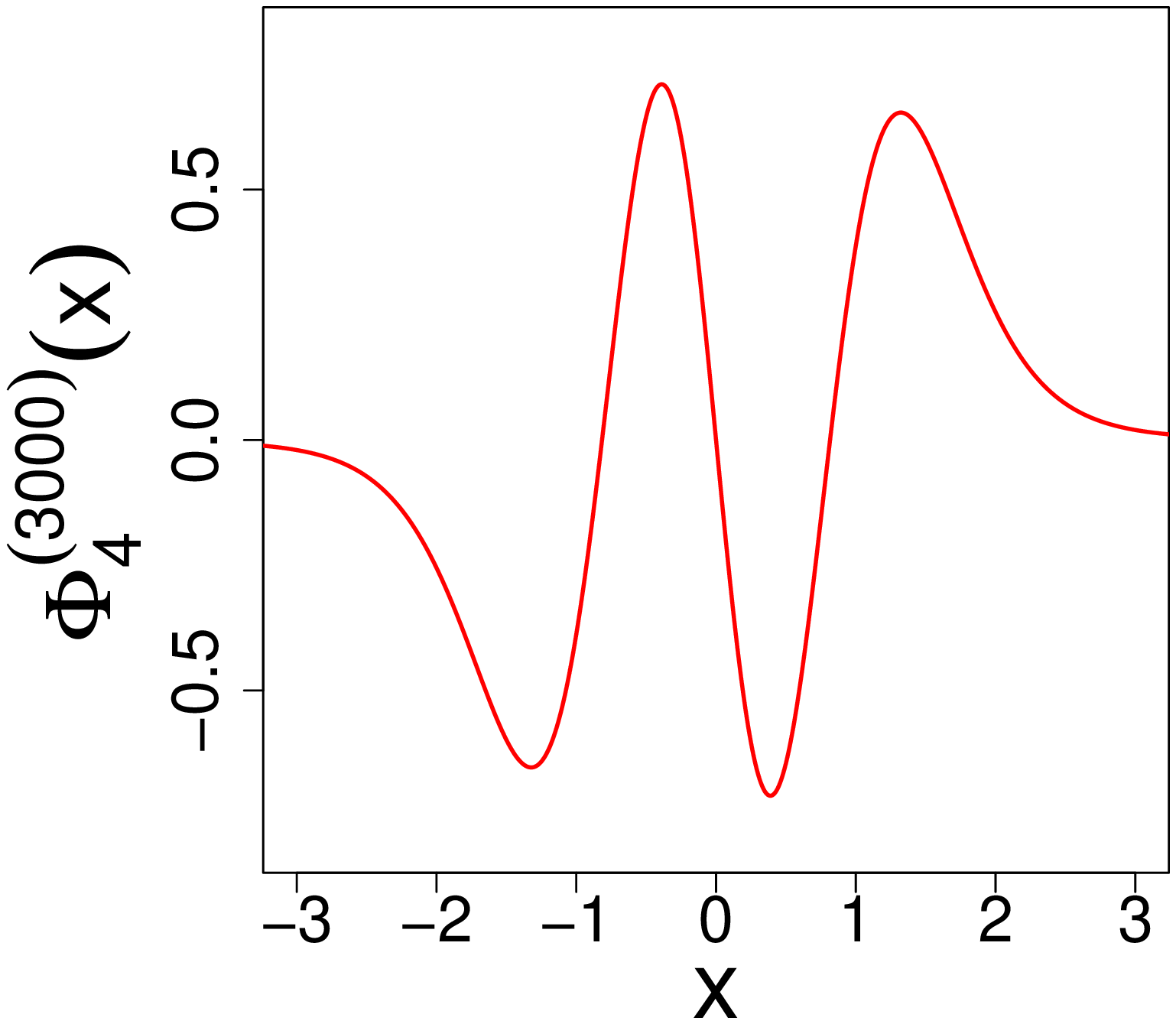}
\includegraphics[width=50mm,height=50mm]{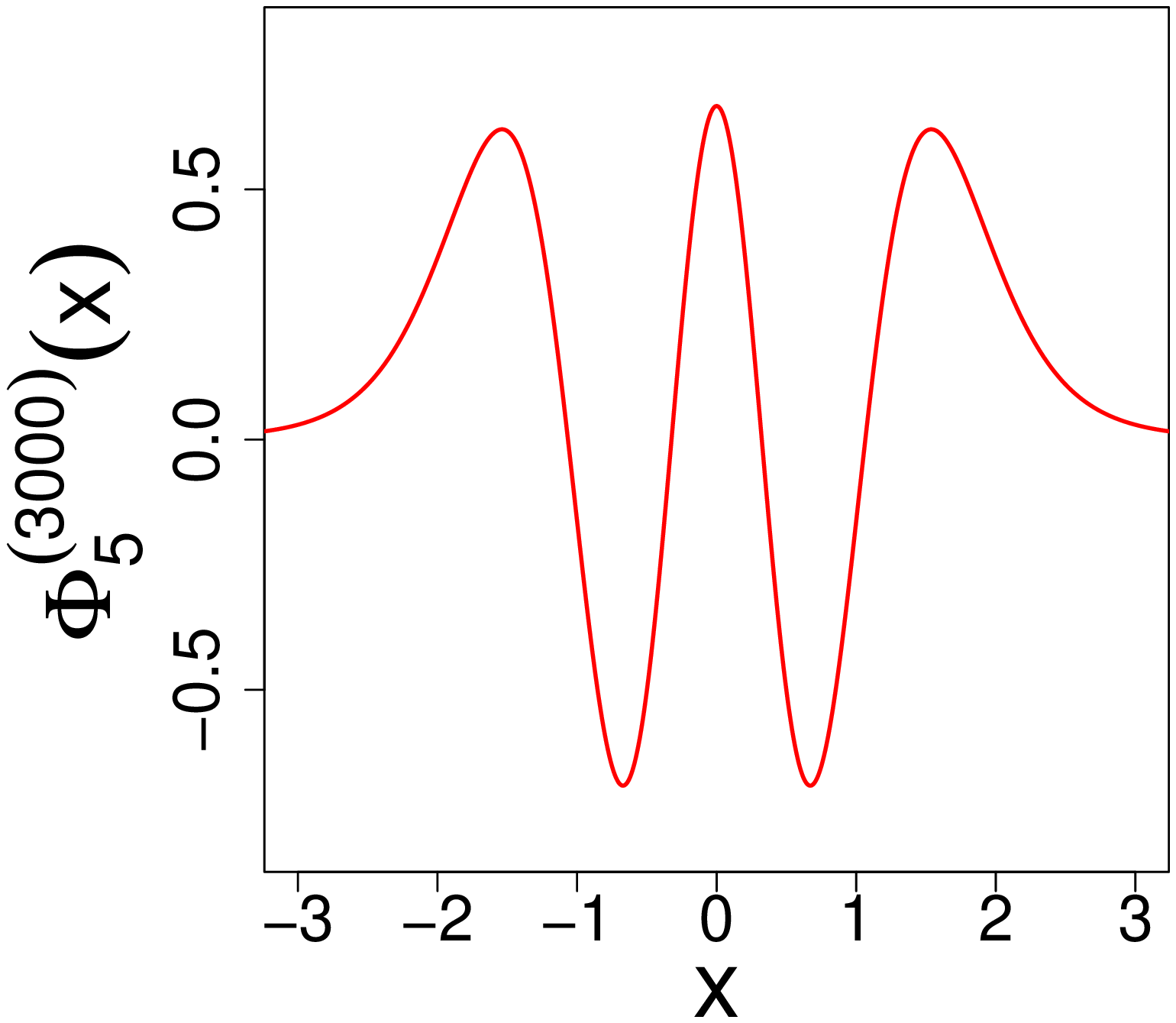}
\caption{The $k$-evolution of $E_i^{(k)}$ for $i=3,4,5$ in the computation with $0\leq i\leq 4$  Hermite trial functions. The dependence on $a$ is displayed as well.
The corresponding (both limiting  and approximating) eigenfunctions, labeled by  $i=3,4,5$,   are depicted  for  $x\in [-3,3]$.   $a\geq 50$ is used
in the course of algorithmic iterations.}
\end{center}
\end{figure}

  If the trial set is composed of    $\Phi_2^{(0)}(x)$ and $\Phi_4^{(0)}(x)$  then   $\psi  _2$ together with  $\psi _4$ will  asymptotically   come out.
 The corresponding eigenvalues read  $E_2=2.33811$  and  $E_4=4.08795$.

 For the algorithm to work  correctly, one needs to   fix    a priori and  next  secure a pre-defined  order  in the Gram - Schmidt  orthonormalization
  procedure, at each $k$-th   iteration. Example:
 in the two-element  set     we  declare  the $(1,3)$   Gram-Schmidt   order. This means that in the ordered pair  $\{\Phi_1^{(0)}(x), \Phi_3^{(0)}(x)\} $  after the first $h$-shift we  normalize
   $\Psi ^{(1)}_1(x)=  S(h)\Phi_1^{(0)}$ and  it is
 $\Psi _3^{(1)}=S(h)\Phi_3^{(0)}$ which   is next (Gram-Schmidt)-transformed to its   ultimate   $\Phi _3^{(1)}$    form (as an  orthonormal complement to
  $\Phi _1^{(1)}$). That is to be continued  until  a terminal  $(1,3)$ ordered  pair  $\{\Psi ^{(k+1)}_1= S(h)\Phi_1^{(k)}, \Psi _3^{(k+1)}=S(h)\Phi_3^{(k)}\}$ is recovered.

In Fig. 5 we  display the $k$ -evolution of  $E_i^{(k)}$, where $i=1, 3$ or $i=3, 1$,  in  direct correspondence with the  above exemplary discussion. We display as well the evolution with the reverse
Gram-Schmidt order ansatz. Here $(3,1)$ means that we  first normalize $\Psi _3^{(k)}$  and  next make the   G-S   transformation  of  $\Psi _1^{(k)}$ to a orthonormal complement $\Psi _1^{(k)}$
  of $\Psi _3^{(k)}$.

We have thus handled the ground state and three consecutive excited states up to $\psi _4$.
The procedure in case of higher states, like e.g. $n\geq 5$,  seems to   need  a bit more  general approach. Namely,
  to get a $(k)$-convergent  approximation of an odd  eigenfunction   $\psi _5 $,   we should  in principle
    consider a    trial set composed of five elements (three of them are even).
This amounts to forming  an initial  (G-S ordered )  trial  set, which  is composed of  five
consecutive  $0\leq n\leq 4$  Hermite functions.

 Nonetheless, to the  same end, an  optimal guess   is to consider merely  three trial functions, say $\Phi_1^{(0)}, \Phi_3^{(0)},\Phi_5^{(0)} $.
   Outcomes of algorithmic iterations are presented in Fig. 6.
We display the $k$-evolution,  up to $k=3000$,   of  $E_i^{(k)}$  for selected indices $i=3, 4, 5,$  and for   values   $a=50, 100, 200$ of the boundary parameter.
We  also  depict  the  approximations    (\ref{l2})  of   the corresponding   eigenfunctions  $\psi _i, i=3,4,5$   of the Cauchy oscillator,   \cite{SG,LM}.
 We note that the plots (e.g.  shapes)   of approximate eigenfunctions  for  $a\geqslant 50$ are   graphically  {\it fapp } independent of $a$. Their defining properties can be read out by
 restricting the space axis in Fig. 6 to the interval $[-3,3]$ and  ({\it fapp })   appear     to  coincide with those for true eigenfunctions.

As a word of warning, let us call again our "linear ansatz" in the Strang splitting formula (5).
For higher  eigenfunctions and eigenvalues,   the  G-S  orthonormalization   procedure
 induces an accumulation of systematic (ansatz related)  errors.
   Therefore one should not uncritically  extend our  algorithm  (designed with a purpose  to investigate
    lowest parts of the spectrum)   to higher parts of the spectrum.

\subsection{Spatial nonlocality impact on the  approximation  finesse.}

As mentioned repeatedly above, the  usage of the  algorithm involves a spatial cutoff placing the simulation outcomes in $L^2([-a,a])$,
even if initial trial functions are elements of  $L^2(R)$.  Interestingly, the shapes of eigenfunctions   appear not
 to be very  sensitive upon the value of $a$. Indeed, all relevant numerical outcomes
(see e.g. the   approximate eigenfunctions plots in   Figs. $ 1, 3, 6$) could   have been reproduced in the space interval $[-3,3]$.
Even though $a=50, 100, 200, 500$ is  employed  in all integrations.

The situation appears to be different  if we pass to approximate eigenvalues, where the $a$ - dependence appears to be vital. Then, convergence
 properties of $E_i^{(k)}(a)$ can be quantified by means of  the detuning (deviation) value $\Delta E_i(a)$ which falls down with the growth
 of $a$,  as displayed in  Figs. $2$ and $4$.

We shall  derive  a rough estimate of the detuning dependence on $a$,  based on the assumption that {\it far} beyond
 the boundaries  of the  interval  $[-3,3]$,
the approximate and true  (limiting) eigenfunction, both  corresponding to the true eigenvalue $E$,  do not differ considerably.

 Let $H(a)f  \sim E(a) f$  sets  an  approximate eigenvalue  solution of the  original Cauchy oscillator problem.
The "kinetic" term   $T(a) f$  has the form dictated by  Eq. (2),  provided   the Cauchy principal value of the  integral
 is evaluated in the  interval $[-a, a]$, $a=50$,
instead of the previously mentioned $[-3,3]$.    Accordingly  for $x\in [-a,a]$  we have
 \be
 H(a) f (x)\sim
   [(-\Delta)^{1/2}   +V(x)]\,f(x)=\frac{1}{\pi}\int_{|z|\leq a} \frac{f(x)- f(x+z)}{z^2}dz  +V(x)f(x) \sim E(a)\, f(x).
\ee
 Let $g(x)$ be another  approximate spectral solution  corresponding  to the eigenvalue $E(b)$, with $b>a$,  that is close to $E(a)$  (as a member of a
  sequence converging to the limiting/genuine  eigenvalue $E$ associated with an eigenfunction $\psi (x)$).

  By our $f\sim g$ assumption, a continuation of $f$ from $[-a,a]$ to  $[-b,b]$  does not differ considerably from $g$ well beyond the control
  interval  $[-3,3]$ (c.f. Figs. 1, 3, 6). We  additionally assume the same about $V(x)g(x)$ (for $V(x) \sim x^2$ this would need
   $g(x) \ll 1/x^2$, in consistency with the estimate  (11)).
        Therefore for $x\in [-b,b]$ we have (keeping in mind that  we evaluate the Cauchy principal value of the integral):
  \be
H(b) g(x)\sim   { \frac{1}{\pi}} \int_{|z|\leq b} {\frac{g(x)- g(x+z)}{z^2}}  dz  +V(x)g(x) \sim E(a)\, g(x)  +
\frac{1}{\pi}\int_{a\leq |z|\leq b} \frac{g(x)- g(x+z)}{z^2}dz  \sim
\ee
$$
 E(a)\, g(x)  +   {\frac{2}{\pi}}\, g(x)  \,  \int_{a}^{b}  {\frac{dz}{z^2}} \sim
\left[ E(a) + {\frac{2}{\pi }}\left( {\frac{1}{a}} - {\frac{1}{b}}\right) \right]\, g(x) \sim E(b) g(x)
 $$
Hence
\be
E(b) -  E(a) \sim  {\frac{2}{\pi }}\left( {\frac{1}{a}} - {\frac{1}{b}}\right).
\ee
Inserting   consecutive  boundary values    $50, 100, 200, 500$  in the above formula we arrive at  a perfect agreement with the
 numerically retrieved  $E(b)- E(a)$   data.
  Namely, we have: $(2/\pi )(1/50 - 1/100)\sim 0.0064$,   $(2/\pi )(1/100 - 1/200)\sim 0.0032$,  $(2/\pi )(1/200 - 1/500)\sim 0.0019$  and  ultimately
  $(2/\pi )(1/500  - 1/\infty )\sim 0.0013$.

   We  can readily  reproduce  the detuning $\Delta E(a)=E-E(a)$  dependence on $a$, as depicted in    Figs. $1$ through $6$, provided the $k$-instant of   the
  numerically implemented evolution is common for all $E(a)$.

\section{Cauchy well eigenvalue problem}

A primary motivation for the present research have been inconsistencies and erroneous statements   detected in  publications devoted to selected fractional QM  spectral problems,
the harmonic potential  and  the  infinite well  problem being included.
Spectral solutions  for the infinite well presented in Refs. \cite{DX, B, D} are incorrect. See e.g. \cite{YL,J,Z,GS} for a discussion of some of those issues, specifically in
connection with a correct spatial shape of the pertinent eigenfunctions.

We take a constructive attitude and instead of  reproducing a list of faulty statements, we shall directly address the finite and next  infinite well problem
in  the  fractional QM, for an exemplary Cauchy case.   The main methodology will based on the previously tested  algorithm,
but our approach will directly   refer   to the finite well  in the  nonlocal (fractional)  context.
Subsequently (like in \cite{GK} where a similar route has been followed in the  local QM)  we shall  pass from a
shallow to a very deep (eventually infinitely deep) well.

Since  approximate spectral formulas and eigenfunction estimates are available
for the the "fractional Laplace operator in the interval" (mathematicians' transcript of  the infinite well  in fractional QM), \cite{K},
the latter publication   will be our reference point as far as an approximation issue  of an   infinite   well in terms of a sequence of  deepening finite wells,
 will become    our concern.

If we choose in Eqs. (1)  and  (2)  the potential $V(x)$  in the form
\be
V(x)=\left\{
       \begin{array}{ll}
         0, & \hbox{$|x|<1$;} \\
         V_0, & \hbox{$|x|\geqslant 1$.}
       \end{array}
     \right.\label{l5}
\ee
where $V_0>0$, we tell about the Cauchy well spectral problem.

In case of  the Cauchy well,  our numerical  algorithm  (i)-(v)  allows to  deduce  approximate eigenvalues and eigenfunctions of the  pertinent  spectral problem.
 Since an  infinite well limit, to which we continually refer, can be formulated in terms of $L^2([-a,a])$ (see however \cite{GK} for another viewpoint in the local context),
 we quite intentionally choose
  an initially pre-defined  trial set of linearly
independent functions  to be an orthonormal basis  in
 $L^2[-1,1]$ whose trivial  extension  to $L^2(R)$  reads  as follows:
\be
\Phi_{n=2m+1}^{(0)}(x)=\left\{
                    \begin{array}{ll}
                      A\cos\left(\frac{n\pi x}{2}\right), & \hbox{$|x|<1$,} \\
                      0, & \hbox{$|x|\geqslant 1$}
                    \end{array}
                  \right.\qquad
\Phi_{n=2m}^{(0)}(x)=\left\{
                    \begin{array}{ll}
                      A\sin\left(\frac{n\pi x}{2}\right), & \hbox{$|x|<1$,} \\
                      0, & \hbox{$|x|\geqslant 1$}
                    \end{array}
                  \right.\qquad
m=0,1,\ldots
\ee
The normalization constant $A$  equals  $\pm 1$. A particular sign choice has no physical meaning, but to compare our simulation outcomes with results of other publications,
sometimes a concrete sign adjustment is necessary (to be explicitly stated when necessary). \\

{\bf Remark 5:}  The above  orthonormal basis in $L^2([-1,1])$  has been claimed in Refs. \cite{DX, B, D}
 to comprise  eigenfunctions  of the infinite Cauchy well problem with boundaries at  the ends of  $[-1,1]$  and the  eigenvalues  $E_n = n\pi /2$.
   These statements we shall openly  disprove  in below, see e.f. Fig. 7,  and  as a constructive part of the argument we shall   provide  numerically  deduced,
 actual  eigenvalues and  shapes of the respective eigenfunctions. The level of accuracy for low lying
 eigenstates is surprisingly good.
It is worthwhile to mention that the only eigenvalue formulas  for an infinite Cauchy well,   that can be termed   rigorous,
refer to the large $n$ limit. More explicitly \cite{K}  we actually have  an approximate expression for $E_n$, provided $n$ is sufficiently large:
\be
 E_n = {\frac{n\pi }2}  - {\frac{\pi}8} + O\left({\frac{1}{n}}\right). \nonumber
 \ee
For completeness of arguments,  let us  give an explicit expression for approximate eigenfuctions
 associated with the large $n$ part of the infinite Cauchy   well spectrum.
Namely, we have (with minor adjustments of the original notation  of Ref. \cite{K}):
\be
\psi_n(x)= q(-x)F_n(1+x)-(-1)^nq(x)F_n(1-x),\qquad x\in R,\nonumber
\ee
where  $E_n=\frac{n\pi}{2}-\frac{\pi}{8}$ and  $q(x)$   is an auxiliary function
\be
q(x)=\left\{
       \begin{array}{ll}
         0 & \hbox{for $x\in(-\infty,-\frac{1}{3})$,} \\
         \frac{9}{2}\left(x+\frac{1}{3}\right)^2 & \hbox{for $x\in(-\frac{1}{3},0)$,} \\
         1-\frac{9}{2}\left(x-\frac{1}{3}\right)^2 & \hbox{for $x\in(0,\frac{1}{3})$,} \\
         1 & \hbox{for $x\in(\frac{1}{3},\infty)$.}
       \end{array}
     \right.\nonumber
\ee
The function $F_n(x)$  is defined as   follows: $
F_n(x)=\sin\left( E_n\,  x+\frac{\pi}{8}\right)-G(E_n \,  x)$,
where  $G(x)$  is the Laplace transform $
G(x)=\int\limits_0^\infty e^{-x s}\gamma(s) ds $ of a positive definite  function $\gamma(s)$:
\be
\gamma(s)=\frac{1}{\pi\sqrt{2}}\frac{s}{1+s^2}\exp\left(-\frac{1}{\pi}\int\limits_0^\infty\frac{1}{1+r^2}\log(1+r s)dr\right).\nonumber
\ee
Evidently, things are here much more  complicated than  an oversimplified (and faulty)  guess (16)
of previous authors would suggest.

\begin{figure}[h]
\begin{center}
\centering
\includegraphics[width=55mm,height=55mm]{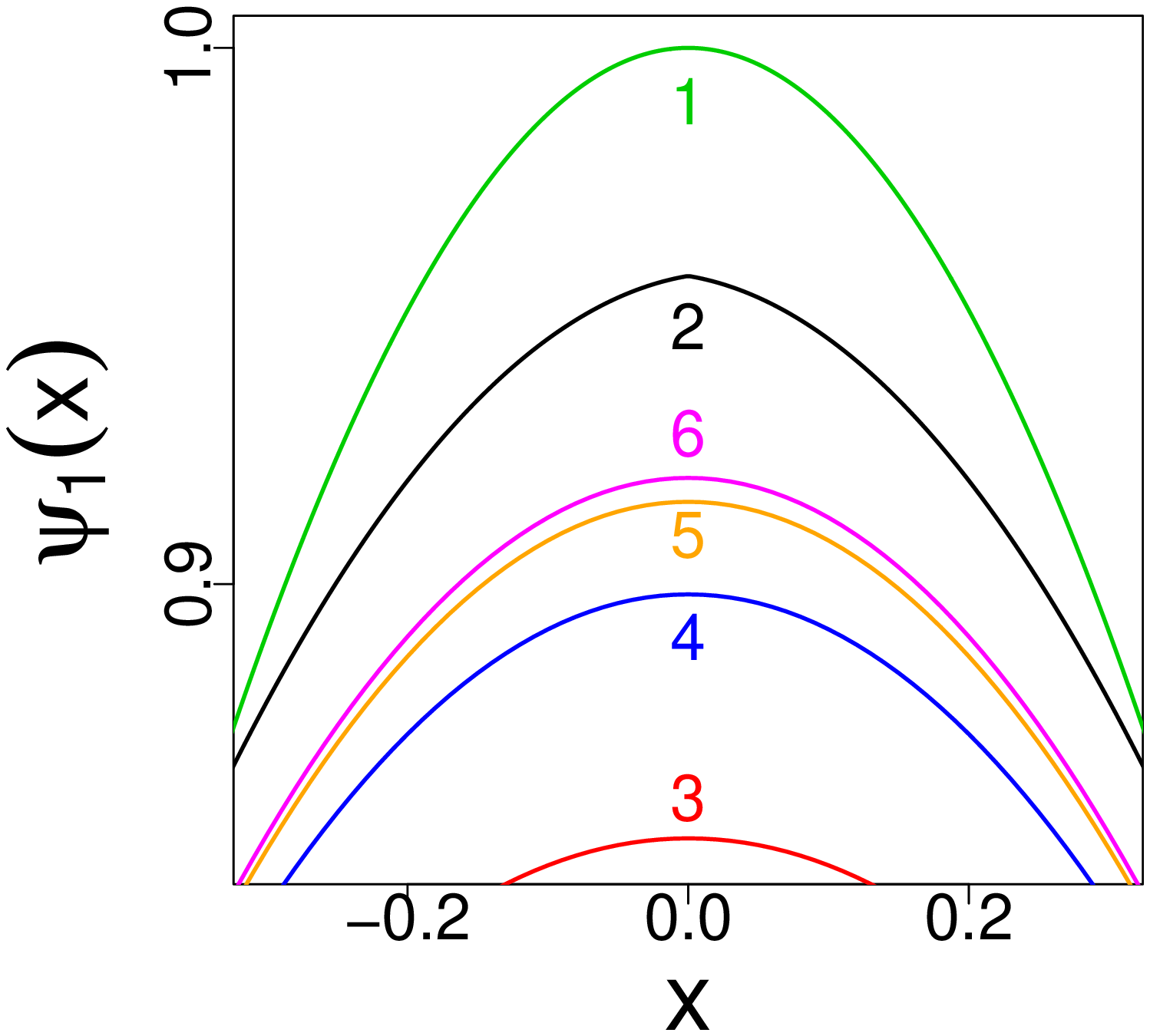}
\includegraphics[width=55mm,height=55mm]{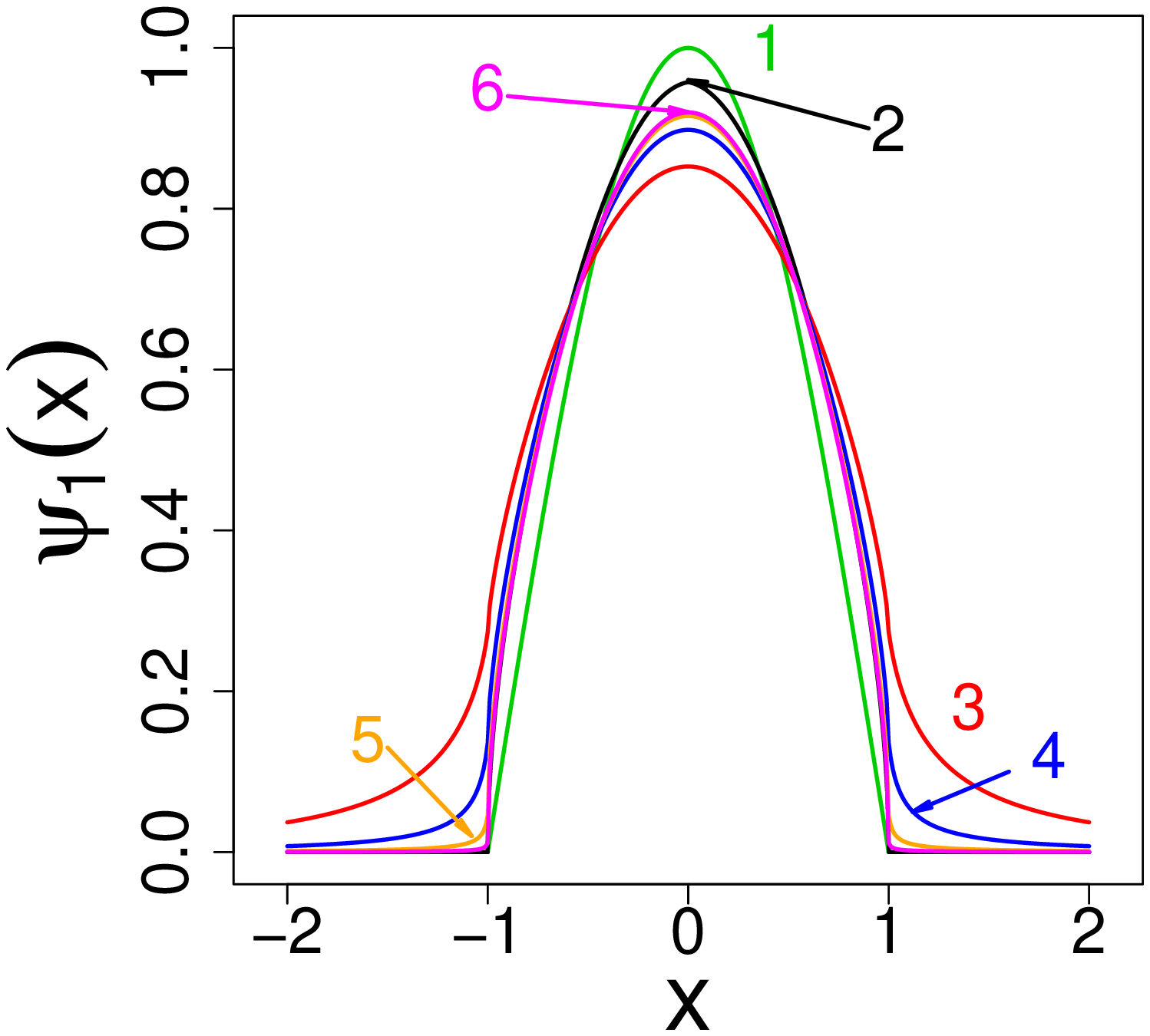}
\includegraphics[width=55mm,height=55mm]{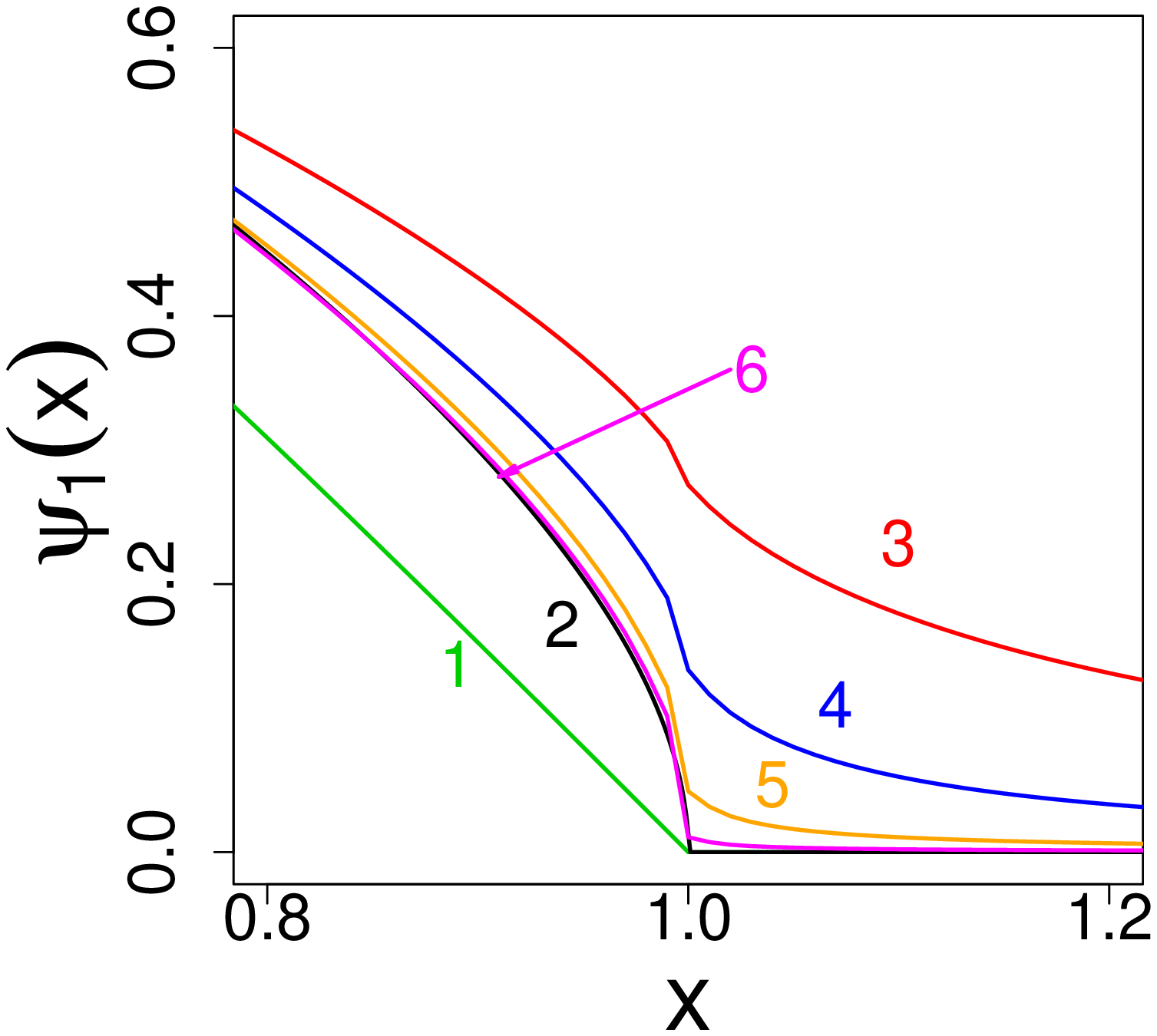}
\caption{Ground state solution of  the Cauchy  well. Numbers refer to: 1 - $\cos(\pi x/2)$, 2 - an approximate solution, Eq. (13) in \cite{K}, 3,4,5,6
  refer to the  well depths, respectively  $5, 20, 100, 500$. Convergence symptoms (towards an infinite well solution) are visually identifiable.
Left  panel reproduces an  enlarged  resolution around the maximum of the  ground state. The right panel does the same job  in the vicinity of
the right boundary  $+1$  of the well (curves deformation comes from scales used to increase a resolution).}
\end{center}
\end{figure}
\begin{figure}[h]
\begin{center}
\centering
\includegraphics[width=58mm,height=58mm]{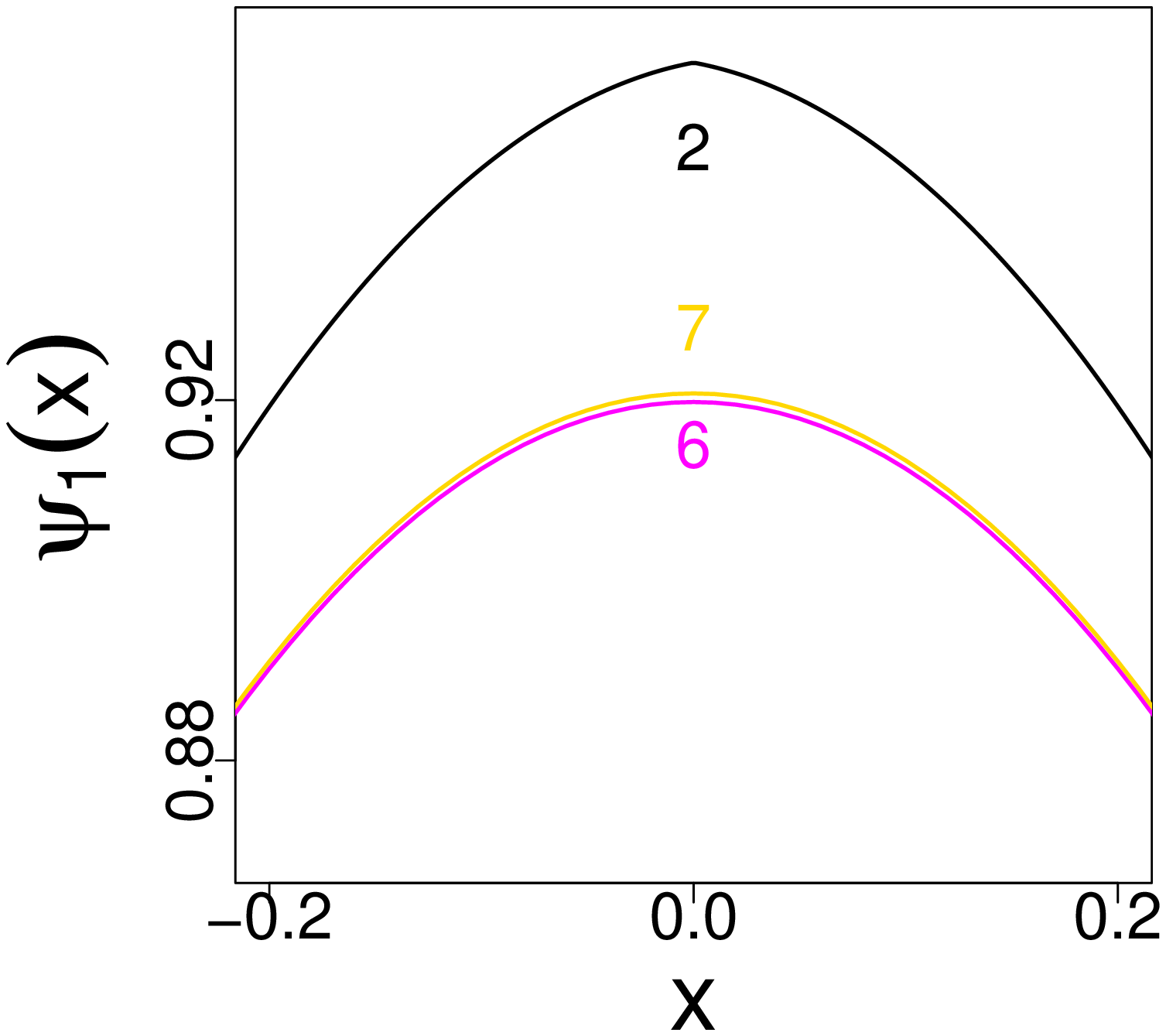}
\includegraphics[width=58mm,height=58mm]{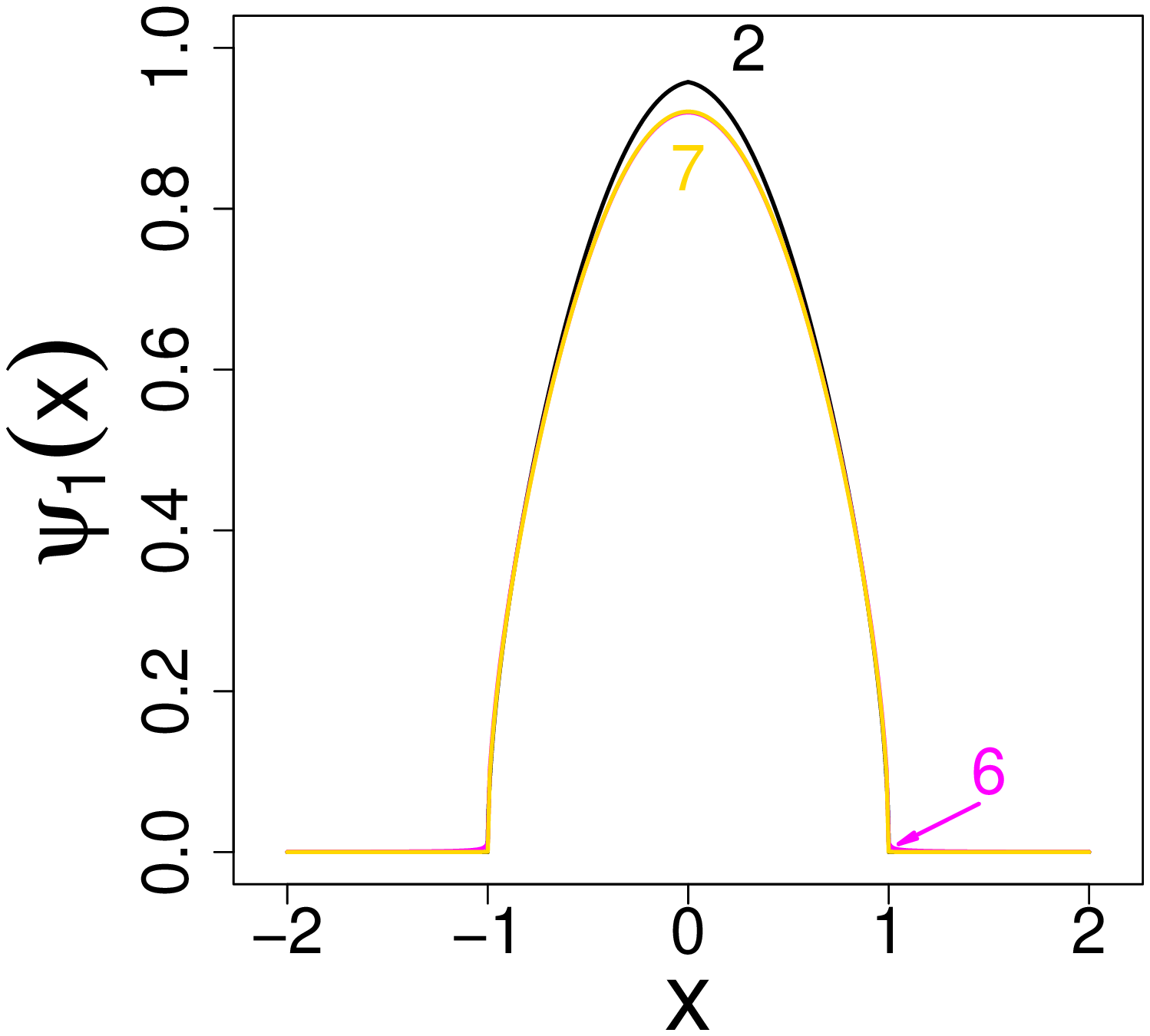}
\includegraphics[width=58mm,height=58mm]{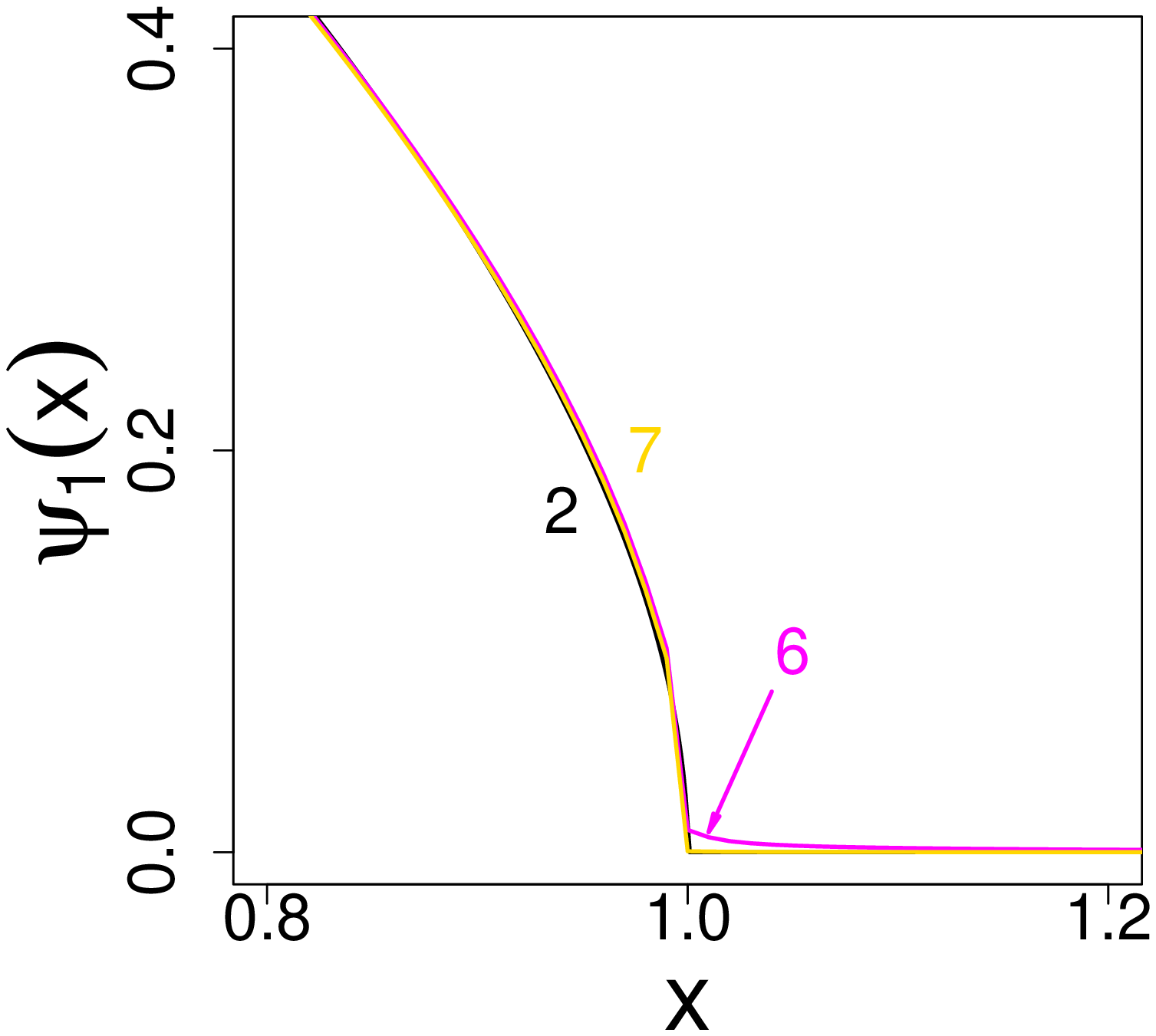}
\caption{Convergence towards $\psi _1$: 2 -  an approximate  ground state, Eq. (13) in  \cite{K}; Our algorithm appears to be more reliable, since  $6$ and $7$  refer
to wells whose depths are respectively $500$ and $5000$. Left panel  shows an enlarged  vicinity of the maxima.  Right panel shows enlarged plots in the vicinity of $+1$.}
\end{center}
\end{figure}

   \begin{figure}[h]
   \begin{center}
   \centering
   \includegraphics[width=95mm,height=95mm]{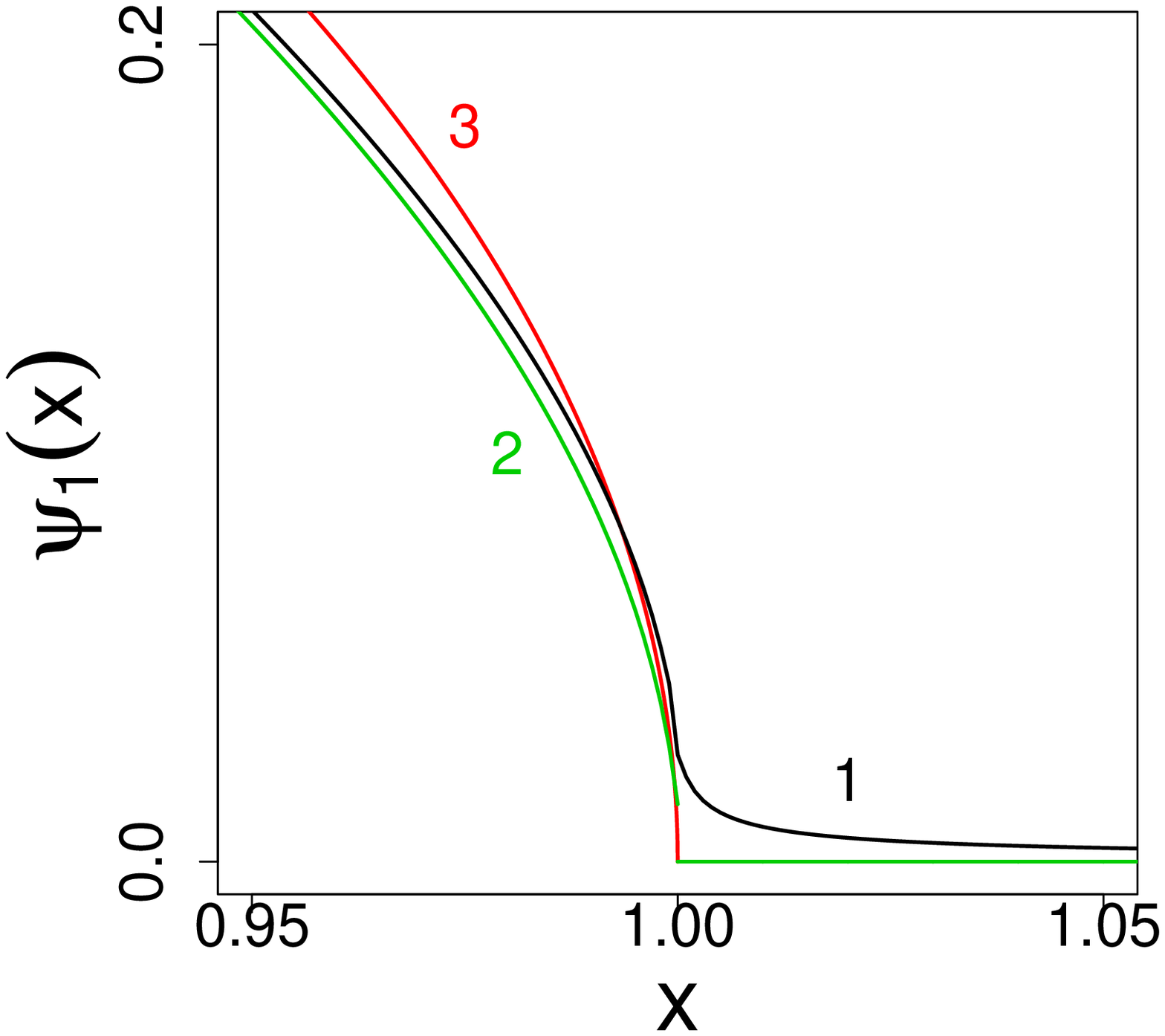}
   \caption{Finite versus  infinite  Cauchy  well ground state  in the vicinity of  the boundary $+1$ of $[-1,1]$: 1 (black) - the  algorithm   outcome for the finite  $V_0=500$  well,
    2 (green)  - an  approximate infinite well  expression from Ref. \cite{K}, 3 (red) -  an approximate form  of
   $\psi _1(x)\sim (1-|x|)^{1/2}$, in the vicinity of the  infinite well barriers,  as proposed in  Ref. \cite{Z}.}
   \end{center}
   \end{figure}

\subsection{Ground state}

As the initial trial function  in $L^2(R)$ we take   $\Phi_{1}^{(0)}(x) = cos(\pi x/2)$  for $|x|< 1$ and $0$ otherwise, see Eq. (16).
Without further mentions of the  detailed (i)-(v)  algorithm steps (c.f.  Sections II.B and C), we shall immediately pass to a discussion of results obtained by means of numerical procedures.
We note that the action of $S(h)$,  Eq. (6),  defined in $L^2(R)$,   effectively   now  takes us away from $L^2([-1,1])$  trial functions to elements of  $L^2([-a,a])$, where $a>1$ sets
actual  integration boundaries. Typically we have    $a\geq 50$.

In  Fig. 7  approximations of the finite well ground state $\psi _1$  are depicted  for few well depths. Namely:
 $V_0=5$ (red, $3$), $V_0=20$ (blue, $4$),  $V_0=100$ (orange, $5$), $V_0=500$ (pink, $6$). These outcomes need to be compared with an apparently faulty  solution  proposed in
  Refs \cite{DX,B,D}(green, $1$) and an approximate ground state  in the infinite well  as  provided in Ref. \cite{K} (black, $2$). \
   Left panel amplifies the differences in the vicinity of a maximum of $\psi _1$, while the right panel
 amplifies them in the vicinity of the right boundary  of the well (e.g. about $+1$). Chosen scales slightly deform the actual  shapes  of  curves, but that is the
  price paid for an  amplification of  differences between closely packed curves.

   We have clearly confirmed (see curves $3$, $4$, $5$ and $6$) a convergence  towards  an infinite well  ground state    $\psi _1$.
Effectively it is   the curve $6$ corresponding  to $V_0= 500$ that suffices to identify  a proper shape  and  a spatial location of the curve
representing the infinite well  ground state $\psi _1$.

  To  get a  better insight into  the convergence issue, in Fig. $8$  we directly compare
approximate ground states that were derived numerically for  $V_0=500$  (pink, $6$)  and  $V_0=5000$ (yellow,  $7$). For comparison,  an  approximate curve drawn  on the basis of
  \cite{K}  is displayed.  Left panel refers to a vicinity of a maximum of $\psi _1$, while the right one  to the  vicinity of the  well boundary.

For finite wells,  quite in affinity with the standard local spectral problem,   eigenfunctions  "spread out" well beyond the well area $[-1,1]$, especially for shallow wells.
Fig. 7 decisively   demonstrates that  $\cos(\pi x/2)$   is not a spectral solution (e.g. ground state)   for an infinite  well problem  (see e.g. curve  1).  It is the vicinity
 of the curves 5 and 6, where the  true  infinite well  ground state is actually located. Compare e.g. Fig. 8 where the depth of the well has been lifted from $500$ to $5000$.
An approximate solution  \cite{K} (curve  2 in Fg. 7) differs considerably  from a true ground state in the central area of the well  while at the well boundaries, together with
all  other  curves appears to converge to a true ground state for  $V_0>100$ (comparison curves are numbered 5 and 6).

  {\bf     Remark 6:}  For  $V_0=5000$  our computer algorithm  appears to  accumulate errors coming from  the  low-order Strang splitting choice and  small-valued  cutoffs finesse.
  That appears not to be significant for the shape of eigenfunctions.   In Fig. 8 we note quite indicative agreement at the boundaries   with the approximating
   curve of \cite{K}.  Somewhat unfortunately, the accuracy  becomes  significantly worse when  it comes to the computation of  eigenvalues, especially for $n>3$. That issue will
    be discussed later.\\

The well  spatial extension is limited to $[-1,1]$.   It seems instructive to know  what the concrete values  (at prescribed points) of eigenfunctions  are beyond this interval.
  This indicates how fast is the decay of the ground state beyond $[-1,1]$.
In Table I we have collected   ground state values at  points $x=  2,10,40, 50$ for various  choices of $V_0$.

\begin{table}[h]
\begin{center}
\begin{tabular}{|l||c|c|c|c|}
\hline
\backslashbox{$V_0$}{$x$} & 2 & 10 & 40 & 50  \\
\hline
\hline
5 & $3.7\cdot 10^{-2}$ & $1.3\cdot 10^{-3}$ & $8.3 \cdot 10^{-5}$ & $5.4\cdot 10^{-5}$ \\
\hline
20 & $7.3\cdot 10^{-3}$ & $2.4\cdot 10^{-4}$ & $1.5 \cdot 10^{-5}$ & $9.6\cdot 10^{-6}$ \\
\hline
100 & $1.3\cdot 10^{-3}$ & $4.4\cdot 10^{-5}$ & $2.7 \cdot 10^{-6}$ & $1.7\cdot 10^{-6}$ \\
\hline
500 & $2.6\cdot 10^{-4}$ & $8.6\cdot 10^{-6}$ & $5.3 \cdot 10^{-7}$ & $3.4\cdot 10^{-7}$ \\
\hline
\end{tabular}
\end{center}
\caption{Values of $\psi _1(x)$ for  $x=2, 10, 40, 50$ and  $V_0= 5, 20, 100, 500$.}
\end{table}

We note a fast convergence of  $\psi _1(x)$  to  $0$, outside of $[-1,1]$. This confirms an antcipated
 $V_0\to\infty$ outcome that $\psi _1(x)$    equals identically zero beyond $(-1,1)$.
  The pertinent limit should be understood in the Cauchy sense:  for  each  $\varepsilon>0$ there exists
   $\delta>0$  such that  for all  $x\in(-\infty,-1]\cup[1,\infty)$,
   if   only $V_0>\delta$,  there holds $|\Phi_1^{(k)}(x)|<\varepsilon$.\\

In Table II,   asymptotic (approximate) finite well eigenvalues are displayed for
$V_0= 5, 20, 100, 500, 5000$ and  integration intervals set by $a=50, 100, 200, 500$.

\begin{table}[h]
\begin{center}
\begin{tabular}{|l||c|c|c|c||c|}
\hline
\backslashbox{a}{$V_0$} & 5 & 20 & 100 & 500 & 5000\\
\hline
\hline
50 & 0.9538 & 1.0743 & 1.1258 & 1.1408 & 1.1445\\
\hline
100 & 0.9602 & 1.0807 & 1.1322 & 1.1472 & 1.1509\\
\hline
200 & 0.9634 & 1.0839 & 1.1353 & 1.1504 & 1.1541\\
\hline
500 & 0.9653 & 1.0858 & 1.1372 & 1.1523 & 1.1560\\
\hline
\end{tabular}
\end{center}
\caption{Approximate ground  state eigenvalue for various well depths  $V_0$  and integration
volume bounds   $a$. The data for $V_0=5000$ were computed with the time step h=0.001.}
\end{table}

We note that the variability of eigenvalues as functions of $a$ does  not depend on the well depth.
Clearly $E(100) - E(50) \sim 0.0064$, $E(200) -  E(100) \sim 0.0032$, $E(500)-E(200) \sim 0.0019$.
These data stay in a precise coincidence with the previously  discussed  effect of extending integration
boundary well beyond the (small)  control interval in the Cauchy oscillator case,  see e.g. (13) and (14).

As directly seen in Table II, the approximate ground state    eigenvalues grow up with the well depth. Tht is not unexpected.
The largest value we  are sure is sufficiently accurate reads  $1.1523$  and corresponds to  $V_0=500$.\\
 The value corresponding to $V_0=5000$  is less accurate, in view of errors  accumulated in the numerical procedure.
 The well depth is so large that we can directly compare this value with that for an infinite well produced in
  Ref. \cite{K}, see Table 2 there in.
  To improve  the fidelity level  for $V_0=5000$, we  need the time partition unit to be one order less  (i.e. h= 0.0001)
   than actually chosen h=0.001.  This decision  would significantly increase the simulation time.

\subsection{Low-lying excited states}

 \begin{figure}[h]
 \begin{center}
 \centering
 \includegraphics[width=58mm,height=58mm]{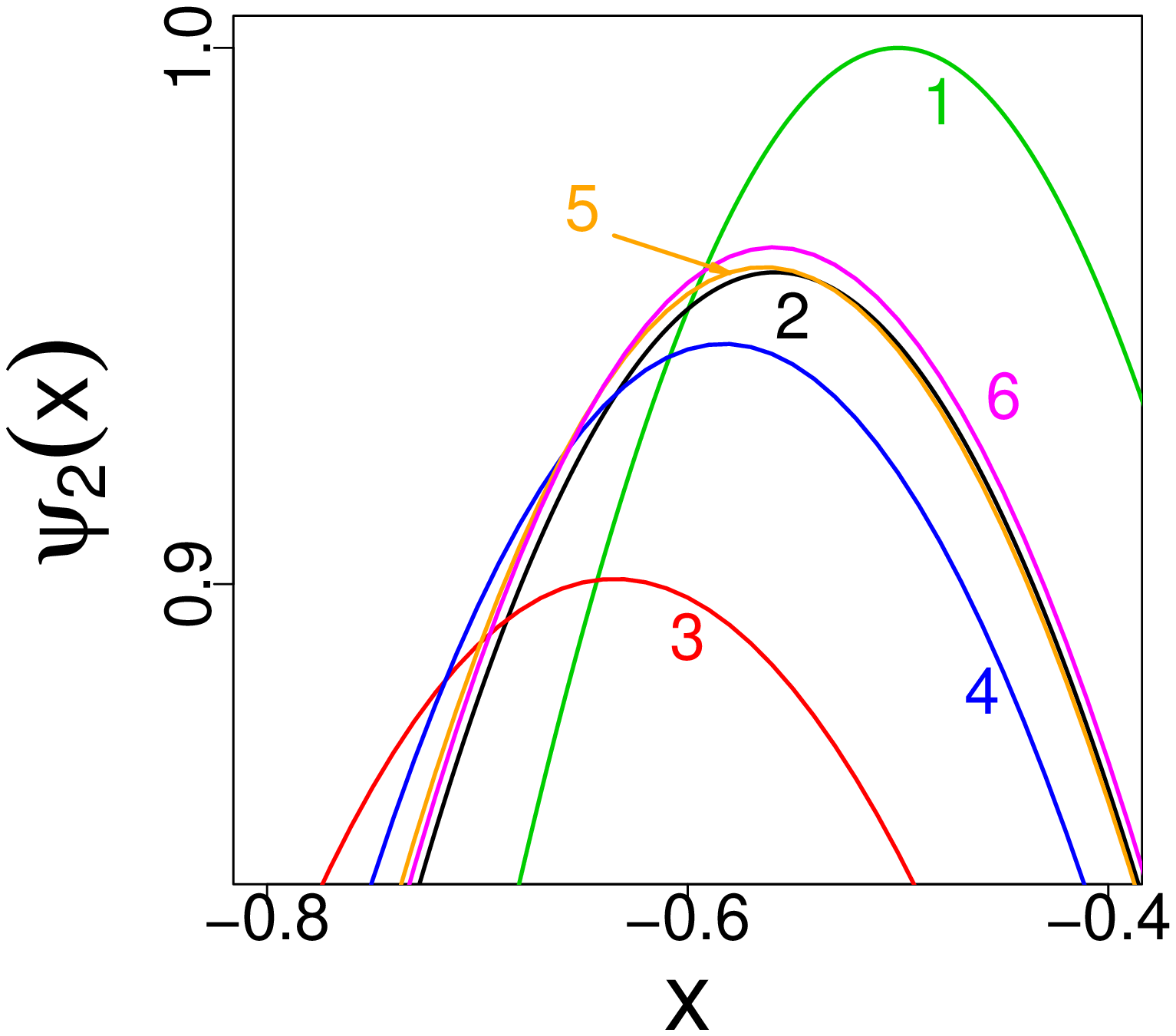}
 \includegraphics[width=58mm,height=58mm]{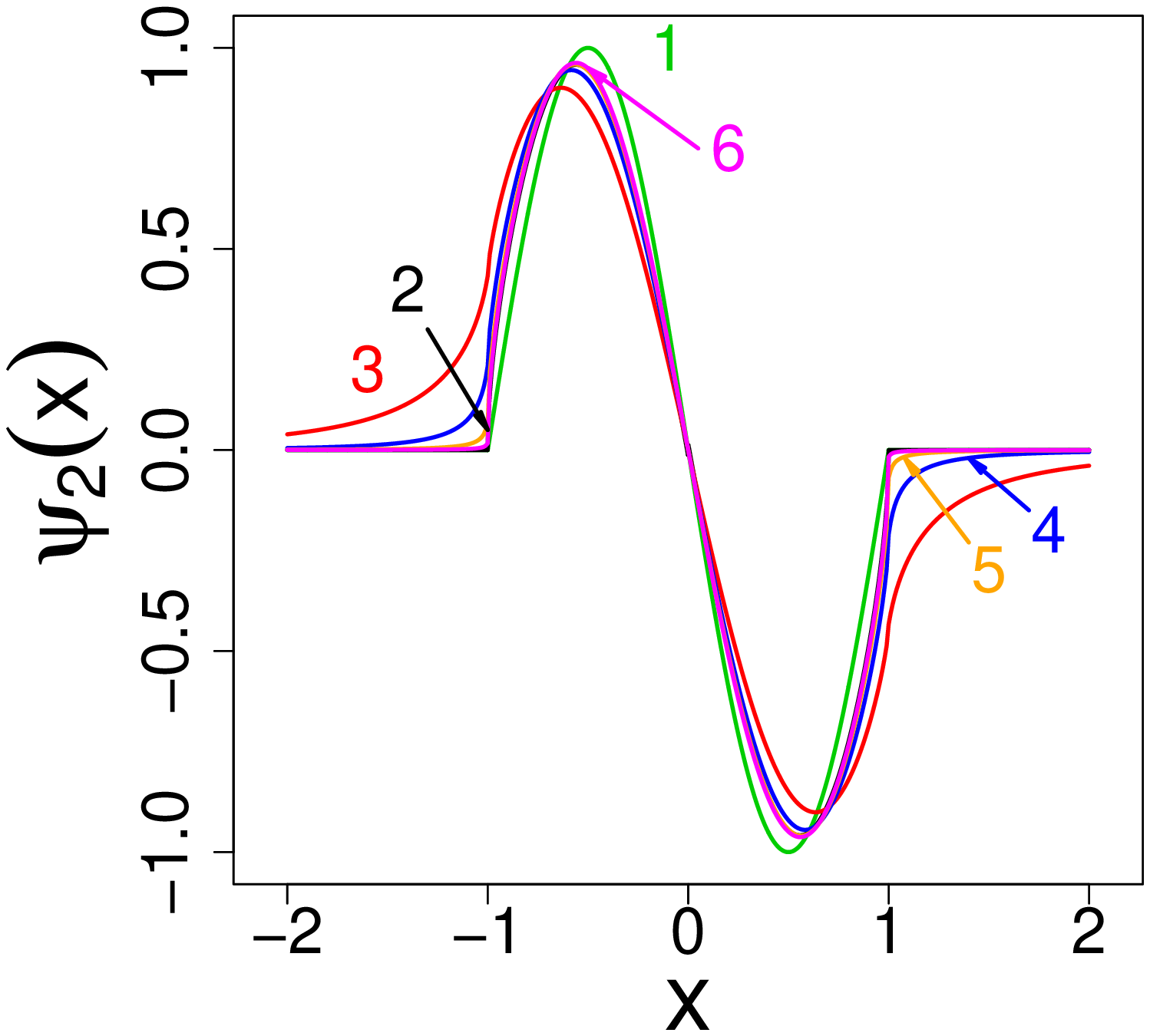}
 \includegraphics[width=58mm,height=58mm]{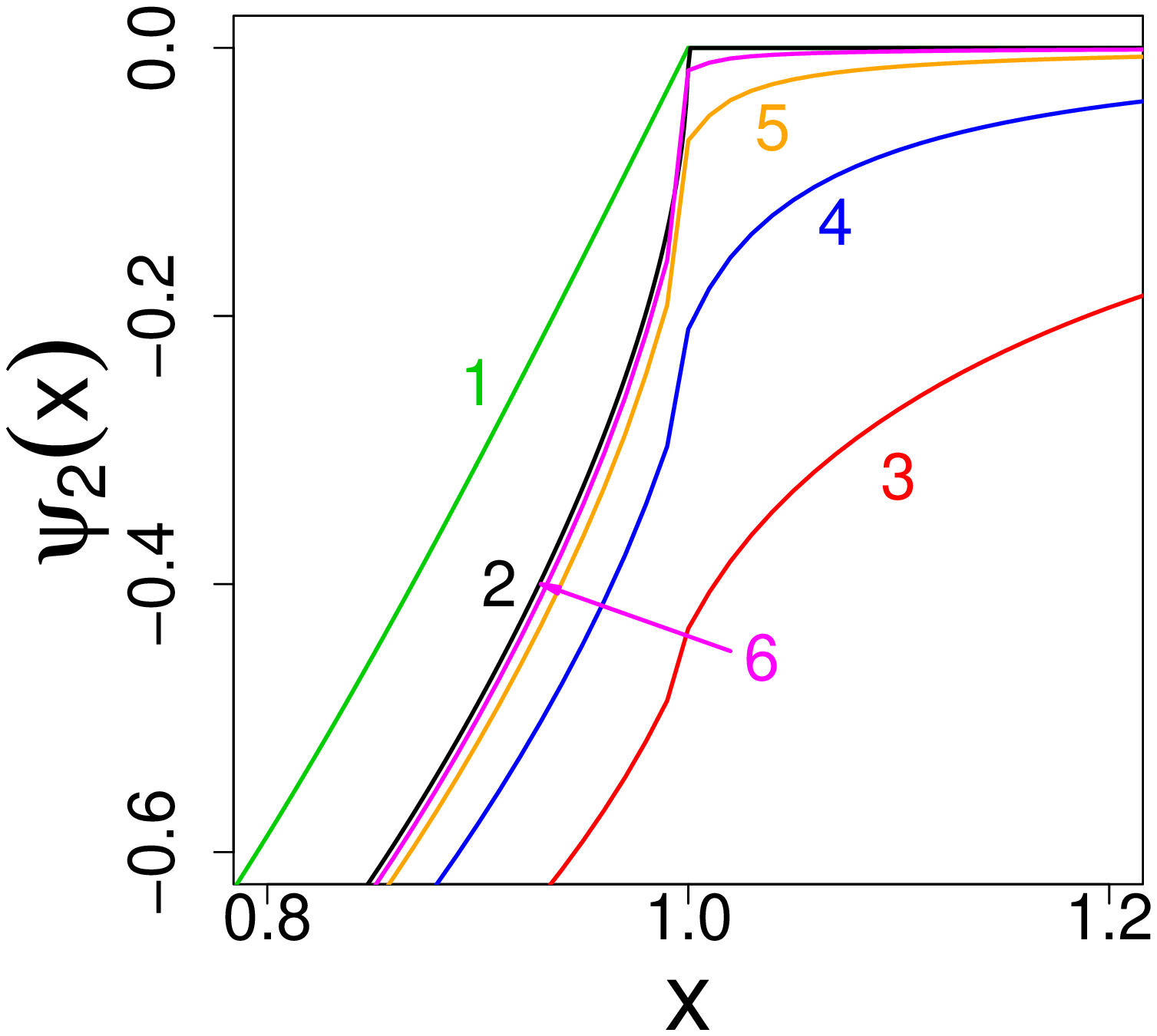}
 \caption{First excited state $\psi _2$. Numbers refer to:  1 - $-\sin(\pi x)$, 2 -  an approximate solution  (13) of Ref.  \cite{K},  3, 4, 5, 6 correspond to finite wells
with depths $V_0=  5, 20, 100, 500$ respectively.  In the left panel we display an enlarged vicinity  of the maximum of $\psi _2$. The right panel contains an enlargement of
  closely packed curves in the vicinity of the right boundary  $1$  of the well.  Note the scales employed. }
 \end{center}
 \end{figure}

\begin{table}[h]
\begin{center}
\begin{tabular}{|l||c|c|c|c||c|}
\hline
\backslashbox{a}{$V_0$} & 5 & 20 & 100 & 500 & 5000\\
\hline
\hline
50 & 2.3701 & 2.6060 & 2.7046 & 2.7343 & 2.7419\\
\hline
100 & 2.3765 & 2.6124 & 2.7110 & 2.7407 & 2.7483\\
\hline
200 & 2.3797 & 2.6156 & 2.7142 & 2.7439 & 2.7515\\
\hline
500 & 2.3816 & 2.6175 & 2.7161 & 2.7458 & 2.7534\\
\hline
\end{tabular}
\end{center}
\caption{The approximate eigenvalue $E_2$ for various well depths  $V_0$  and  $a$.}
\end{table}

To   deduce the first excited level of the finite well (if in existence, we have checked that the shallow well with  $V_0=5$  has three bound states), we  take
$ \Phi_{1}^{(0)}(x)=  -\sin ( \pi x)$ for {$|x|<1$  and $0$ otherwise, see Eq. (16)   as a trial function.
For comparison with Ref. \cite{K} we have chosen the minus sign  instead of the positive one. In Fig. 10 the first excited state $\psi _1$ has been depicted  (colors, curves
 numbering  being  the same as in the ground state displays in Fig. 7).

We have a clear confirmation that would be infinite well eigenfunctions of Refs. \cite{DX, B, D}  are plainly wrong  (curve 1).
 Our approximate eigenfunctions  show a definite  convergence  towards an asymptotic (true infinite well) eigenfunction, see  e.g. curves 5 and 6.
 An  approximate eigenfunction of Ref. \cite{K} is  much better approximation of a true  excited  eigenstate, than it was in case of the ground state. Nonetheless there are obvious deviations
from the true shape of the eigenstate in the vicinity of the well boundaries  (lets panel, curves  2 and 6 need to be compared).

 We anticipate that with the increase of $V_0$  the obtained   modifications of the shape of the  curve 6,  in the vicinity of its extrema,  would be insignificant.
 The behavior in the vicinity of the well boundaries would be more indicative.   Compare e.g. Fig. 9  depicting  the ground state  of the  "almost"  infnite well.

In Table III we have collected approximate  eigenvalues of $\psi _2 $ for various  well  depths  $V_0$  and integration boundary values  $a$.
Again, we observe that  the variability of eigenvalues as functions of $a$ does  not depend on the well depth.
The data stay in a precise coincidence with the previously  discussed  effect of extending integration
boundary well beyond the (small)  control interval in the Cauchy oscillator case,  see e.g. (13) and (14).
The approximate eigenvalue of $\psi _2 $  grow together with $V_0$.

 The largest numerical outcomes that  correspond to  $V_0=500$  and  $V_0=5000$
can be directly compared with values reported in Table 2 of Ref. \cite {K}.
  As a complement to this discussion,  subsequently (Table  VII) we shall compare  our maximal approximate eigenvalues  with best (optimal) approximations reported in Ref.  \cite{K}.

 \begin{figure}[h]
 \begin{center}
 \centering
 \includegraphics[width=58mm,height=58mm]{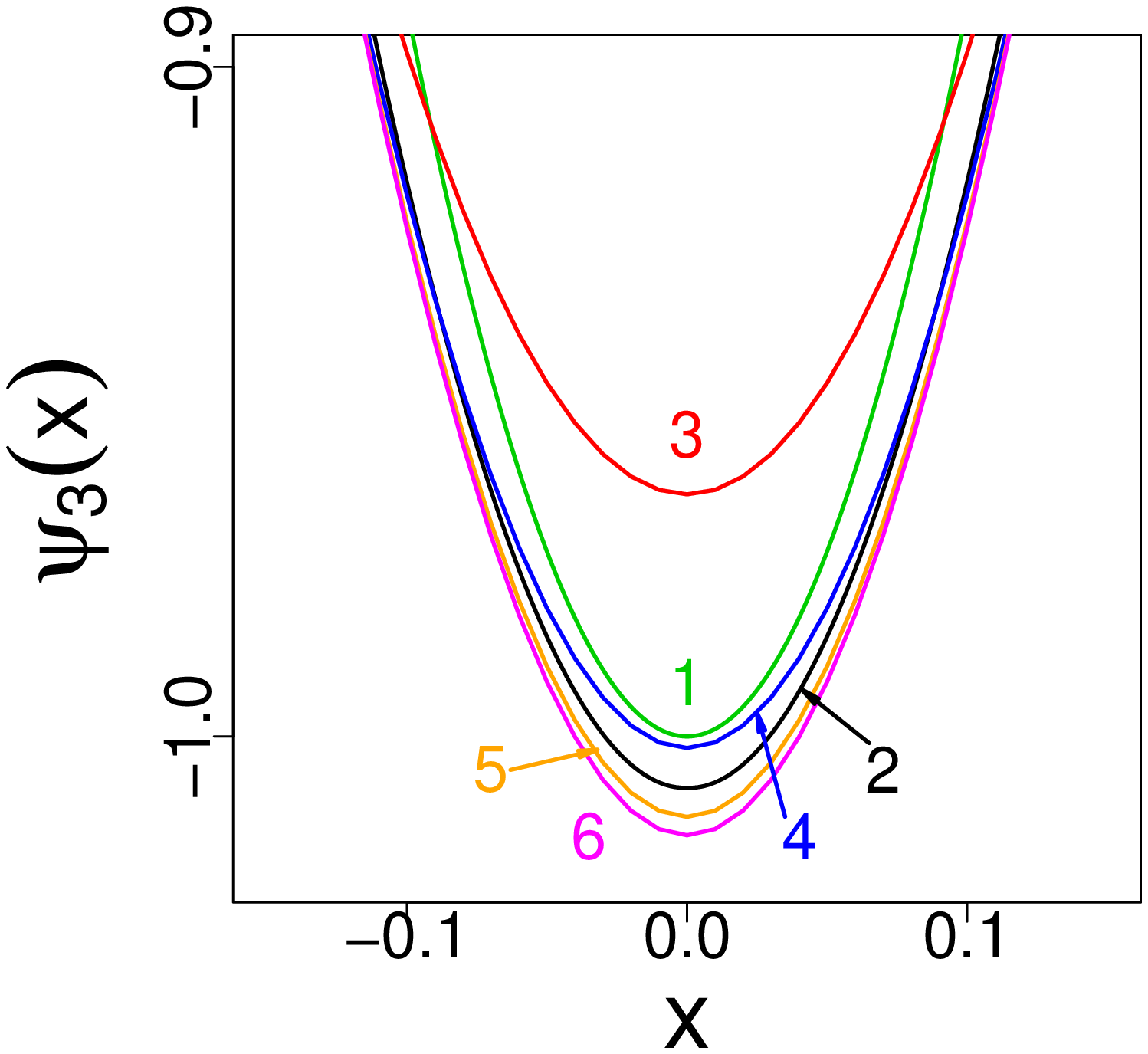}
 \includegraphics[width=58mm,height=58mm]{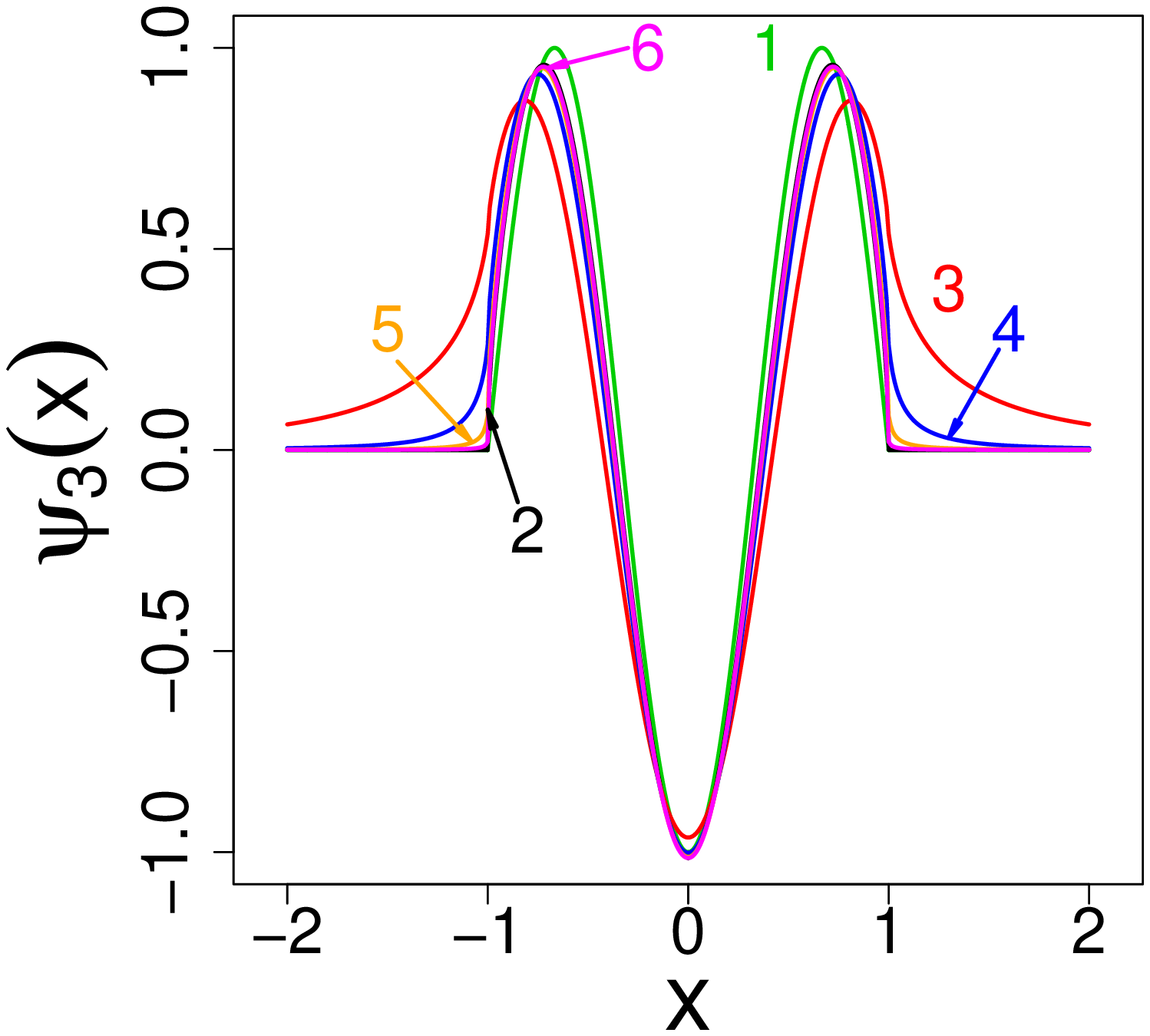}
 \includegraphics[width=58mm,height=58mm]{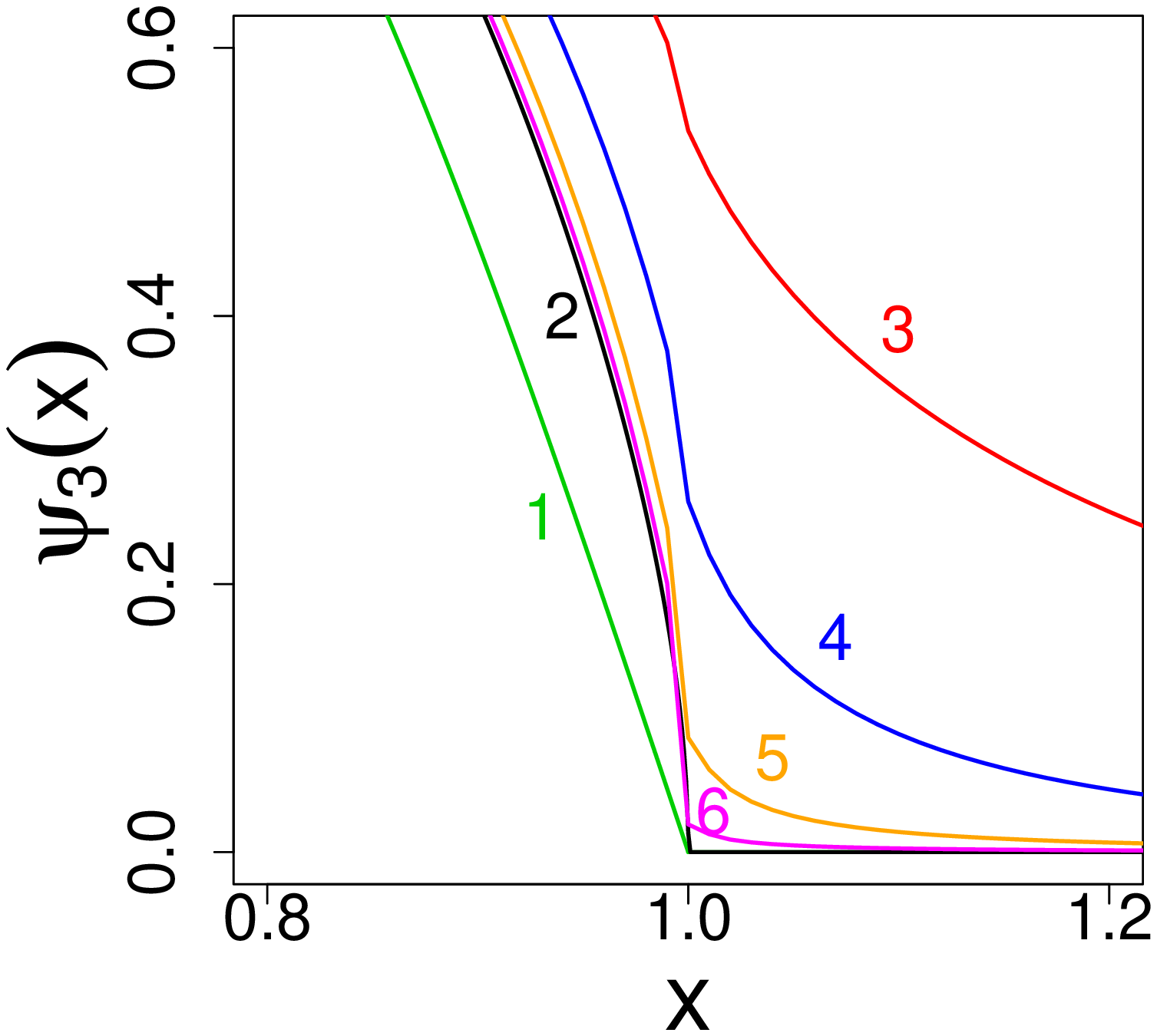}
 \caption{Eigenstate  $\psi _3$. Numbers refer to: 1 - $-\cos(3 \pi x/2)$, 2 -  approximate solution (13) of Ref.  \cite{K}, 3, 4, 5, 6
 correspond to well depths $V_0= 5, 20, 100, 500$.
 Left panel contains an enlargement  of a minimum  of $\psi _3$. Right panel depicts an enlargement of closely packed curves in the vicinity of the well boundary.}
 \end{center}
 \end{figure}
 \begin{figure}[h]
 0\begin{center}
 \centering
 \includegraphics[width=58mm,height=58mm]{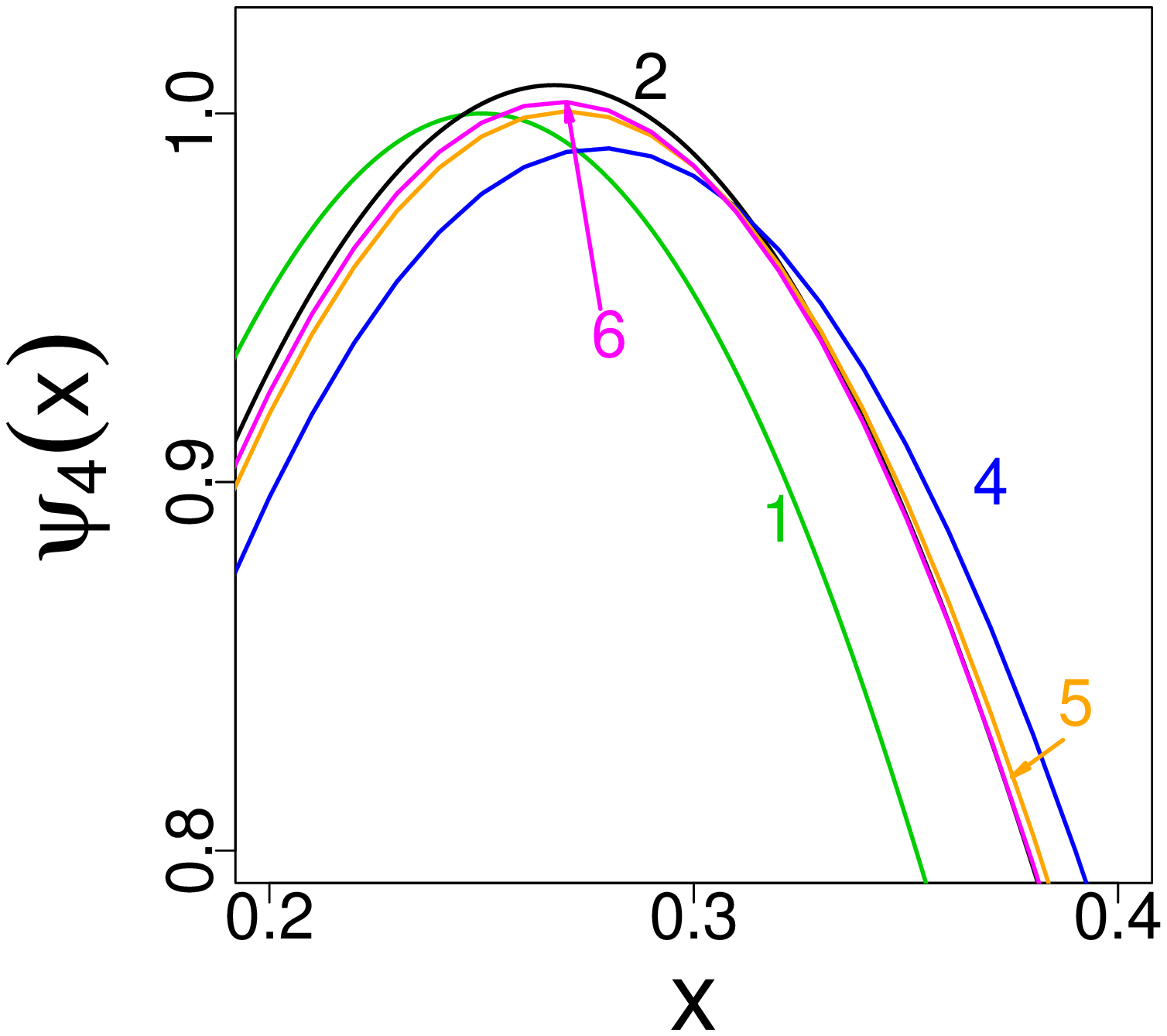}
 \includegraphics[width=58mm,height=58mm]{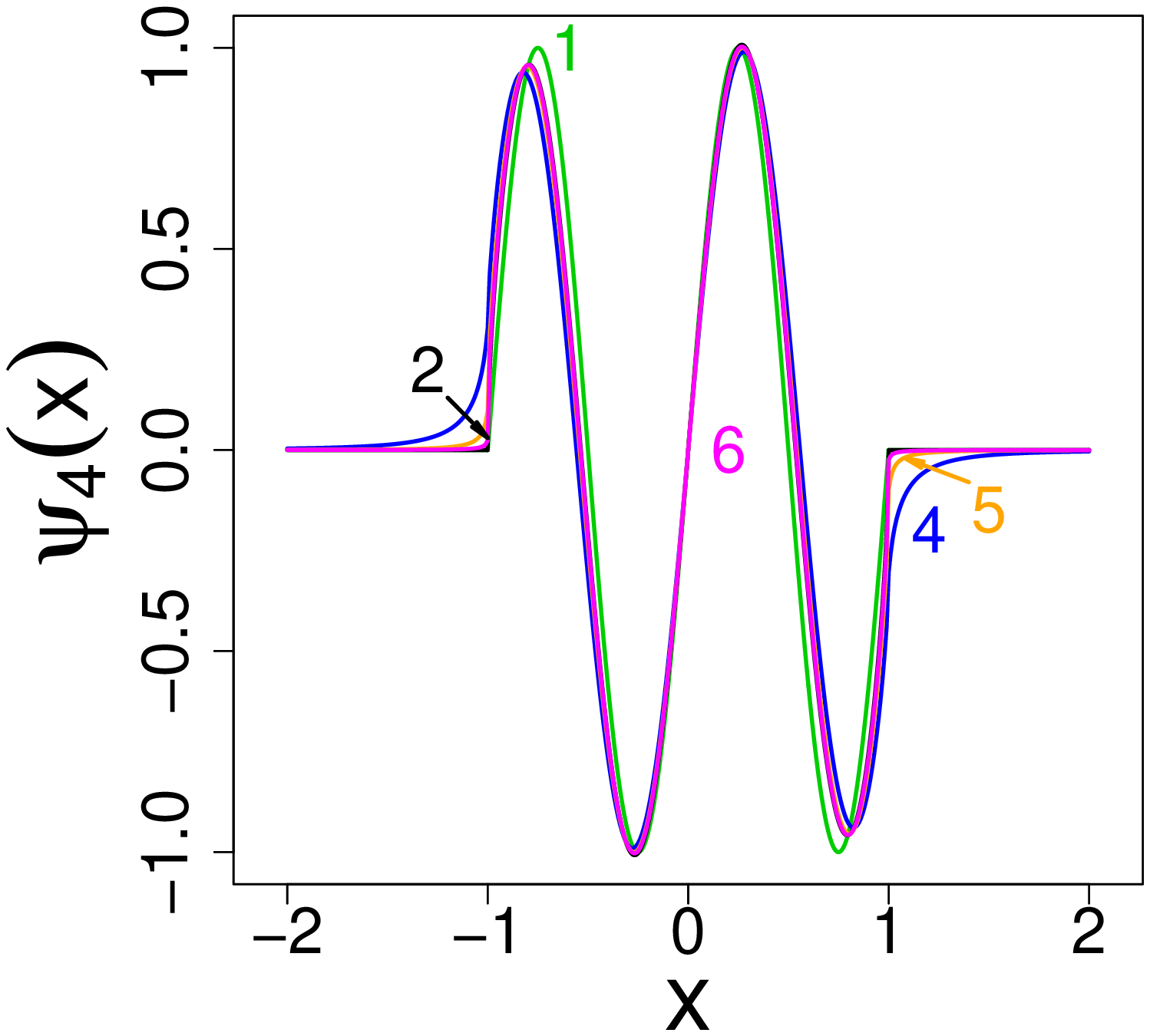}
 \includegraphics[width=58mm,height=58mm]{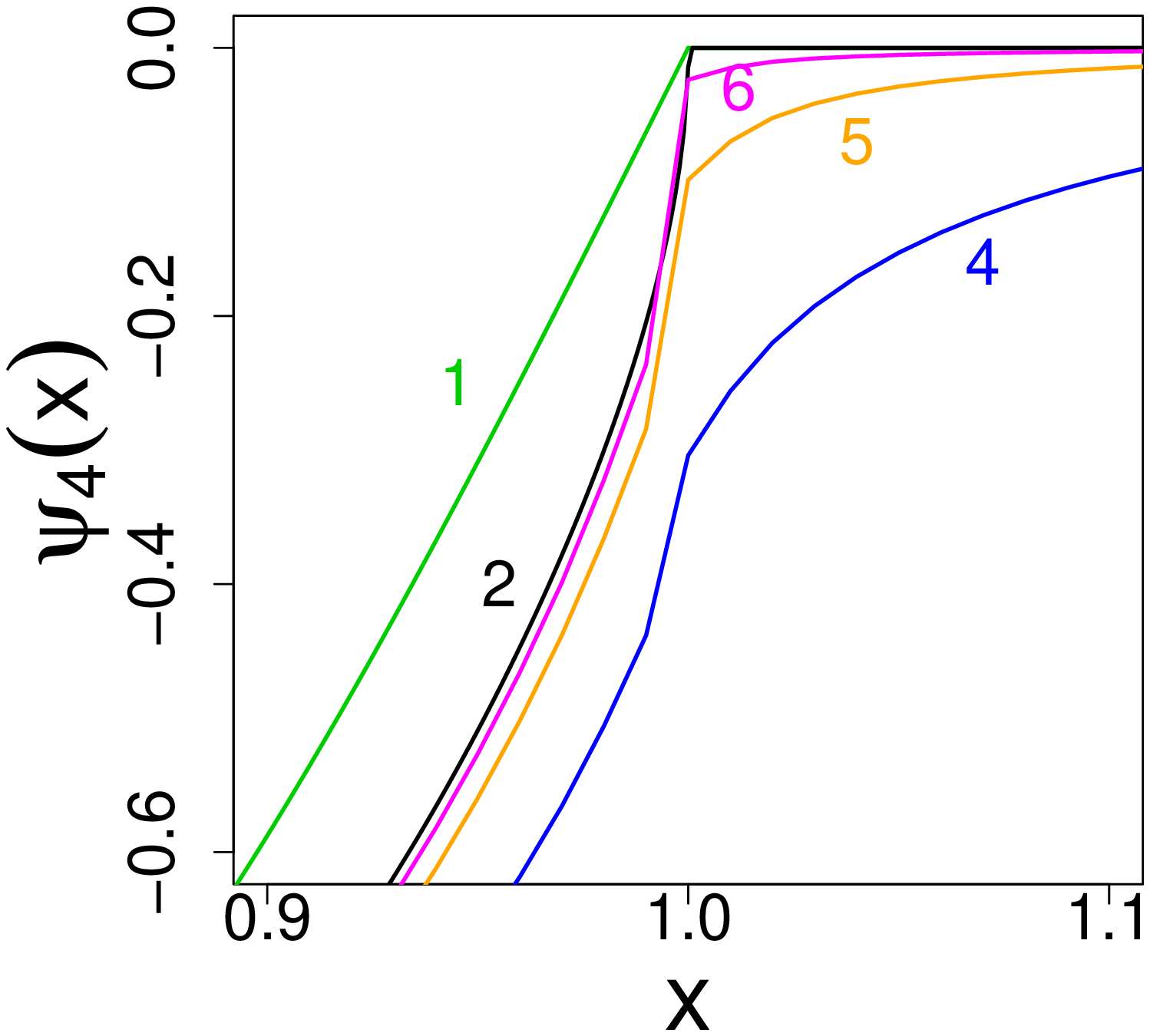}
 \caption{Eigenstate $\psi _4$.  Numbers correspond to: 1 - $\sin(2 \pi x)$, 2 -  approximate solution (13) of Ref.  \cite{K}, 4, 5, 6 refer to $V_0= 20, 100, 500$.
 Left panel depicts the vicinity of a  maximum   of $\psi _4$. Right panel refers to the well boundary.}
 \end{center}
 \end{figure}

\begin{table}[h]
\begin{center}
\begin{tabular}{|l||c|c|c|c||c|}
\hline
\backslashbox{a}{$V_0$} & 5 & 20 & 100 & 500 & 5000\\
\hline
\hline
50 & 3.7457 & 4.1116 & 4.2517 & 4.2942 & 4.3053\\
\hline
100 & 3.7522 & 4.1180 & 4.2581 & 4.3006 & 4.3117\\
\hline
200 & 3.7554 & 4.1212 & 4.2613 & 4.3038 & 4.3149\\
\hline
500 & 3.7573 & 4.1231 & 4.2632 & 4.3057 & 4.3168\\
\hline
\end{tabular}\qquad
\begin{tabular}{|l||c|c|c|c||c|}
\hline
\backslashbox{a}{$V_0$} & 5 & 20 & 100 & 500 & 5000\\
\hline
\hline
50 & - & 5.6378 & 5.8147 & 5.8687 & 5.8830\\
\hline
100 & - & 5.6442 & 5.8211 & 5.8751 & 5.8894\\
\hline
200 & - & 5.6474 & 5.8243 & 5.8783 & 5.8926\\
\hline
500 & - & 5.6493 & 5.8262 & 5.8802 & 5.8945\\
\hline
\end{tabular}
\end{center}
\caption{Approximate values for $E_3$ and $E_4$ for various  $V_0$ and  $a$.
The  $V_0=5$ well has three bound states only.}
\end{table}
\begin{table}[h]
\begin{center}
\begin{tabular}{|l||c|c|c|c||c|}
\hline
\backslashbox{a}{$V_0$} & 5 & 20 & 100 & 500 & 5000\\
\hline
\hline
50 & - & 7.1584 & 7.3720 & 7.4369 & 7.4543\\
\hline
100 & - & 7.1648 & 7.3784 & 7.4433 & 7.4607\\
\hline
200 & - & 7.1680 & 7.3816 & 7.4465 & 7.4639\\
\hline
500 & - & 7.1699 & 7.3835 & 7.4484 & 7.4658\\
\hline
\end{tabular}\qquad
\begin{tabular}{|l||c|c|c|c||c|}
\hline
\backslashbox{a}{$V_0$} & 5 & 20 & 100 & 500 & 5000\\
\hline
\hline
50 & - & 8.6878 & 8.9359 & 9.0109 & 9.0312\\
\hline
100 & - & 8.6942 & 8.9423 & 9.0173 & 9.0376\\
\hline
200 & - & 8.6974 & 8.9455 & 9.0205 & 9.0408\\
\hline
500 & - & 8.6993 & 8.9474 & 9.0224 & 9.0427\\
\hline
\end{tabular}
\end{center}
\caption{Approximate values for $E_5$   and $E_6$. }
\end{table}
\begin{table}[h]
\begin{center}
\begin{tabular}{|l||c|c|c|c||c|}
\hline
\backslashbox{a}{$V_0$} & 5 & 20 & 100 & 500 & 5000\\
\hline
\hline
50 & - & 10.2136 & 10.4979 & 10.5827 & 10.6057\\
\hline
100 & - & 10.2200 & 10.5043 & 10.5891 & 10.6121\\
\hline
200 & - & 10.2232 & 10.5075 & 10.5923 & 10.6153\\
\hline
500 & - & 10.2251 & 10.5094 & 10.5942 & 10.6172\\
\hline
\end{tabular}\qquad
\begin{tabular}{|l||c|c|c|c||c|}
\hline
\backslashbox{a}{$V_0$} & 5 & 20 & 100 & 500 & 5000 \\
\hline
\hline
50 & - & 11.7443 & 12.0638 & 12.1579 & 12.1836\\
\hline
100 & - & 11.7507 & 12.0702 & 12.1643 & 12.1900 \\
\hline
200 & - & 11.7539 & 12.0734 & 12.1675 & 12.1932\\
\hline
500 & - & 11.7558 & 12.0753 & 12.1694 & 12.1951\\
\hline
\end{tabular}
\end{center}
\caption{Approximate eigenvalues for $E_7$ and $E_8$. }
\end{table}
 \begin{table}[h]
  \begin{center}
  \begin{tabular}{|c||c|c|c|c|c|c|c|c|}
  \hline
   & $E_1$ & $E_2$ & $E_3$ & $E_4$ & $E_5$ & $E_6$ & $E_7$ & $E_8$\\
  \hline
  Ref.\cite{K} & 1.1577 & 2.7547 & 4.3168 & 5.8921 & 7.4601 & 9.0328 & 10.6022 & 12.1741\\
  \hline
  $a=500,\,V_0=500$& 1.1523 & 2.7458 & 4.3057 & 5.8802 & 7.4484 & 9.0224 & 10.5942 & 12.1694\\
  \hline
  $a=500,\,V_0=5000$& 1.1560 & 2.7534 & 4.3168 & 5.8945 & 7.4658 & 9.0427 & 10.6172 & 12.1951\\
  \hline
  \end{tabular}
  \end{center}
  \caption{A comparison of our approximate eigenvalues  for a $V_0=500,\, a=500$  with "most optimal" values presented in Ref.  \cite{K}.}
  \end{table}

To deduce   other   excited states and corresponding eigenvalues an algorithm needs to be used in its full extent, with a properly chosen initial set of trial functions and theri consecutive Gram-Schmidt
orthonormalization at each iteration step.  In Figs 11  and 12  we depict  shapes of the  third and fourth  finite well eigenstates (in the well $V_0=5$ three bound states are in existence).
Tables IV-VI collect the data for  six  consecutive eigenvalues (up to $n=8$).  Again, we have clearly confirmed that would-be   infinite well eigenfunctions
of Refs  \cite{DX, B, D} are plainly wrong. Albeit, per force, one can admit some level of robustness, at which those can be viewed as very robust approximations of a true state of affairs.

In Table VII we have compared  maximal simulated eigenvalues, obtained  for  $V_0=500$, $V_0=5000$ and  $a=500$,
with optimal approximations reported in Ref.  \cite{K}.
Our  numerical values  of Table VII, in case of   $V_0=500$  are slightly lower than  those in Table 2 of Ref. \cite{K}, as expected.
Namely, our computed values refer to finite wells, while the data of Ref. \cite{K} refer to an infinite well.
We can  make  a rough guess that the computed value for  $a=500$,  might differ by about $0.0013$
from an infinite well eigenvalue (c.f. our previous $E(a)- E(b)$ discussion).

Remembering about a remark we have made before  about an accumulation of numerical errors when we pass to higher eigenstates and eigenvalues
(they are basically a consequence of the Gram-Schmidt orthonormalization procedure),    we  may look at respective eigenvalues from this "error accumulation" perspective.
A comparison of  our numerical results   in the  $V_0=5000$ case
 with approximate eigenvalues of Ref. \cite{K} indicates  that,  beginning from $n=4$,  our procedure shows up some limitations and needs improvements (those are known for practitioners
 of analogous algorithms in the local case). The resultant eigenvalues become too large, compared with our anticipation.
 The simplest way to improve our  numerics would be to chose so small time step  $h$, such that for larger $V_0$ we would have at least
  $h V_0 <1$, and even better if  $h V_0 <1/2$.

  \section{Outlook}

Our numerically-assisted  derivation of eigenfunctions and eigenvalues may be adopted to any nonlocal problem (e.g. any fractional or quasi-relativistic motion generator), not only in 1D but also
in 2D or 3D.  There is no basic limitation upon the  choice of the external confining potential $V(x)$.   The main model-dependent factors  are the time interval   partition unit $h$ and the size of
integration volume (that was $a$ in our 1D case, while an arbitrary  finite  open volume in $R^3$ is admitted). They  affect the ultimate numerical outcomes.

Coming back to our 1D considerations, an important issue is that of
  seeking an optimum: small time partition unit  $h$ versus large $V_0$}.
In particular, for fairly small $h$, higher Taylor series  terms can be accounted for in the Strang splitting method. It is know that in the local case,
 the  fourth-order algorithm works pretty well \cite{Auer}.

 For the finite well potential  (\ref{l5})  the time step  $h$ needs to be adjusted to $V_0$ so
 that $h\,V_0$ is "small enough". In the present paper we were mostly interested in the precise deduction of lowest eigenstates and
eigenvalues. To this end,  the
choice of  $h=0.001$ and well depths  $V_0=5\,,20\,,100\,,500$  proved to be optimal.

 As we have mentioned before our decision to employ the Gram-Schmidt orthonormalization procedure at each step of the algorithm proved
 to be a source of accumulating errors. Other procedures are known to be more reliable. However, we have been seeking for an optimal choice between various
 error-producing factors and the simulation time. For lowest eigenstates the G-S choice was optimal in this respect.

The major goal of the paper, that of disproving faulty results concerning the infinite well spectrum has been achieved. We  have produced a number of high
accuracy numerically-assisted  approximations of "true"  finite and (ultimately)   infinite Cauchy well   eigenstates. A comparison with approximate results
obtained in the mathematical literature \cite{K},
proved an efficiency of both our computation method and the reliability of its outcomes.

\end{document}